\documentclass[pdflatex,sn-mathphys-num]{sn-jnl}

\usepackage[utf8]{inputenc}
\usepackage[T1]{fontenc}
\usepackage{xcolor}          
\usepackage{bm}
\usepackage{adjustbox}
\usepackage{amssymb}
\usepackage{amsmath}
\usepackage{booktabs}
\usepackage{placeins}
\usepackage{multirow}
\usepackage{graphicx}
\usepackage{array}
\usepackage{ragged2e}
\usepackage{arydshln}        
\usepackage{subfigure}
\usepackage{stfloats}        
\usepackage{floatrow}
\usepackage{tabularx}
\usepackage{makecell}
\usepackage{tablefootnote}
\usepackage{longtable}
\usepackage[linesnumbered,ruled,vlined]{algorithm2e}

\newfloatcommand{capbtabbox}{table}[][\FBwidth]

\title{Data Security in Large Language Models: Risks, Defense, and Directions}

\author[1,2]{Kang Chen}\equalcont{Kang Chen and Xiuze Zhou are the co-first authors.}
\email{chenkang@kean.edu}

\author[3]{Xiuze Zhou}
\email{xz.zhou@connect.hkust-gz.edu.cn}

\author*[1]{Yuanhui Yu}
\email{andy@jmu.edu.cn}

\author*[1]{Yuanguo Lin}
\email{xdlyg@jmu.edu.cn}

\author[1]{Hefeng Chen}
\email{chenhf@jmu.edu.cn}

\author[1]{Congyu Cai}
\email{congyucai@jmu.edu.cn}

\author[4]{Li Shen}
\email{ls6743@nyu.edu}

\affil[1]{School of Computer Engineering, Jimei University, Xiamen 361021, China}
\affil[2]{College of Science, Mathematics and Technology, Wenzhou-Kean University, Wenzhou 325060, China}
\affil[3]{The Hong Kong University of Science and Technology (Guangzhou), Guangzhou, China}
\affil[4]{School of Professional Studies, New York University, New York 10003, United States}

\abstract{
Large Language Models (LLMs), now a foundation in advancing natural language processing, power applications such as text generation, machine translation, and conversational systems. Despite their transformative potential, these models inherently rely on massive amounts of training data, often collected from diverse and uncurated sources, which exposes them to serious data security risks. \textcolor{black}{Harmful or malicious data can compromise model behavior, leading to toxic outputs or hallucinations, while also creating vulnerabilities to data-driven attacks such as prompt injection and data poisoning.} As LLMs continue to be integrated into critical real-world systems, understanding and addressing these data-centric security risks is imperative to safeguard user trust and system reliability. This survey offers a comprehensive overview of the main data security risks facing LLMs and reviews current defense strategies, including adversarial training, data cleaning, output guardrails, Reinforcement Learning from Human Feedback (RLHF), data augmentation, and Retrieval-Augmented Generation (RAG)/agent defenses. Additionally, we categorize and analyze relevant datasets used for assessing robustness and security across different domains, providing guidance for future research. Finally, we highlight key research directions that focus on data provenance and traceability, verifiable machine forgetting, secure model updates, \textcolor{black}{standardized evaluation framework,} explainability-driven security analysis, and effective governance frameworks, aiming to promote the safe and responsible development of LLM technology. This work seeks to inform researchers, practitioners, and policymakers, driving progress toward data security in LLMs.}


\begin{document}
\maketitle

\section*{Introduction}

\textcolor{black}{Large Language Models (LLMs), which have demonstrated strong performance on tasks ranging from free-form text generation and summarization to machine translation and open-domain question answering, represent a major advance in Natural Language Processing (NLP). }
\textcolor{black}{The ability of LLMs to model complex linguistic dependencies and generate coherent, context-aware outputs has resulted in widespread adoption in both academic research and industrial applications. This progress underscores the significance of LLMs as powerful computational tools and important components of modern AI systems. } 
\textcolor{black}{LLMs have also demonstrated strong in-context learning ability \cite{duan2023flocks}.} 
\textcolor{black}{Their growing adoption has further extended their application across diverse NLP tasks \cite{plant2025you}.}

\textcolor{black}{Despite their remarkable strengths, LLMs face a range of data security vulnerabilities. These issues threaten model integrity and the trustworthiness of generated content. Due to their reliance on massive training datasets, LLMs are particularly susceptible to malicious or biased information, which may lead to the generation of inaccurate or inappropriate content.}
This raises serious concerns about potential negative impacts, such as the spread of false information and the reinforcement of harmful stereotypes. By manipulating public opinion, fostering confusion, and advancing detrimental ideologies, the intentional dissemination of misinformation may cause substantial societal harm \cite{pan2023risk}.
\textcolor{black}{These threats operate at different stages: data poisoning and backdoor injection corrupt training data \cite{nazary2025stealthy}, and jailbreaking exploits inference-time prompt vulnerabilities \cite{wang2025your}, illustrating the dual-edged nature of web-scale data ingestion.}
\textcolor{black}{They can manifest at multiple stages in the LLM lifecycle, thereby compromising model outputs and undermining trust.}
Moreover, the lack of transparency in training data provenance further exacerbates these risks.
\textcolor{black}{From a data-centric perspective, this survey examines how these threats exploit the data lifecycle of LLMs, including training, fine-tuning, and inference data.}
Studies have shown that even small amounts of toxic, biased, or copyrighted content in a training set can disproportionately affect model behavior \cite{carlini2023extracting}. With the ever-widening scale of LLMs, ensuring dataset integrity becomes increasingly critical - not only to prevent harmful generations but also to uphold legal and ethical standards. Recent work highlights the urgency of constructing curated and auditable training corpora to mitigate these issues \cite{biderman2023pythia}. Without such safeguards, LLMs remain susceptible to data-centric threats, which can subtly or overtly distort their outputs.
\textcolor{black}{This survey adopts a data-centric perspective on LLM security, focusing on threats where data is the target, vector, or source of risk across training, fine-tuning, inference, retrieval, and agent interaction. Model-centric topics such as gradient leakage, model stealing, and other parameter- or optimization-level attacks are outside the main scope of this work.}

\begin{figure}[htbp]
  \centering
  \includegraphics[width=\textwidth]{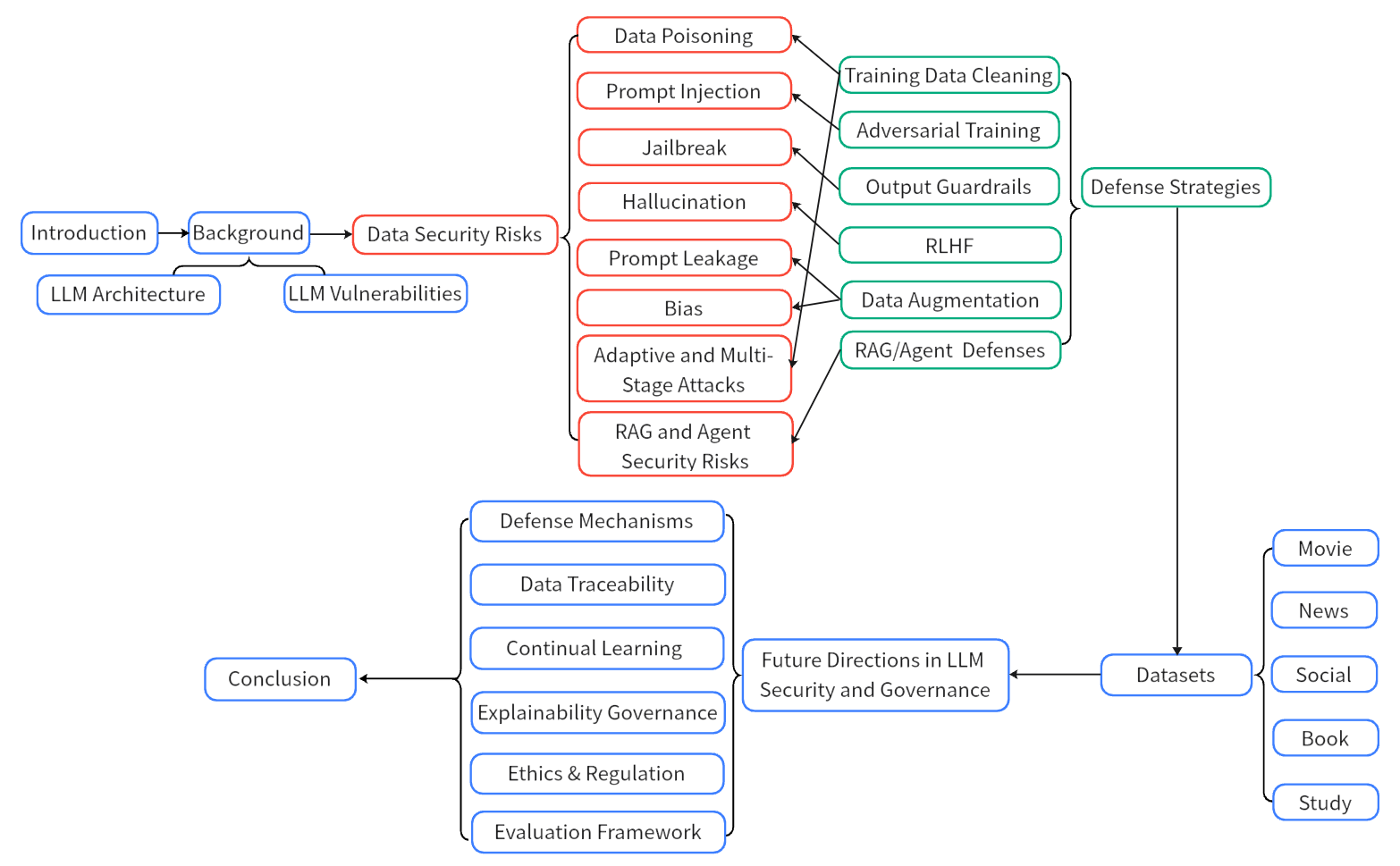}
\caption{\textcolor{black}{Overview of the survey structure on LLM data security. The paper progresses from background and vulnerabilities to data security risks, defense strategies, datasets, and future directions. Arrows indicate the primary defenses surveyed for each risk category (e.g., data cleaning for poisoning, adversarial training and output \textcolor{black}{guardrails} for prompt injection and jailbreak, RLHF and data augmentation for hallucination and bias).}}
  \label{fig0}
\end{figure}

To address these concerns, a range of protective methods has been developed. \textcolor{black}{This survey focuses specifically on data-centric defense mechanisms.} Key data security protection methods include adversarial training \cite{mkadry2017towards}, Reinforcement Learning from Human Feedback (RLHF) \cite{ouyang2022training}, and data augmentation techniques \cite{gallegos2024bias}. 
Recent research increasingly highlights the data security risks associated with training LLMs, particularly their vulnerability to training-time data poisoning. It has been shown that even a small fraction of corrupted training data can significantly undermine model behavior \cite{steinhardt2017certified}. To counter such threats, researchers have proposed robust training frameworks that reduce the impact of manipulated data, aiming to preserve model reliability throughout the learning process \cite{jagielski2018manipulating}. These findings collectively reinforce the importance of embedding security considerations into the entire lifecycle of LLM development.

\textcolor{black}{In addition to these traditional vulnerabilities, \textcolor{black}{from a data-centric security perspective, modern applications that integrate LLMs with Retrieval-Augmented Generation (RAG) can introduce additional data security risks.} While RAG helps ground outputs in external knowledge, it also exposes models to privacy inference attacks, where adversaries can extract sensitive information from document embeddings or vector indices \cite{zeng2024good}. Moreover, corrupted or biased retrieval corpora may manipulate LLM outputs, making RAG systems vulnerable to both content poisoning and unanticipated harmful generations \cite{an2025rag}. Similarly, LLM-based Agent systems, which autonomously perform multi-step reasoning and task delegation, expand the attack surface.} \textcolor{black}{Malicious users can exploit RAG-enabled interactions to extract private datastore content through prompt-injected queries \cite{qi2025follow}.}
\textcolor{black}{In tool-integrated agent workflows, indirect prompt injections embedded in untrusted external content can further coerce agents into executing unintended actions, highlighting a shift from data leakage to operational exploitation \cite{zhan2024injecagent}.} 

Several prior surveys partially explored aspects of data security in LLMs, but often with a narrower scope or focus. Some studies have explored adversarial threats in NLP, offering extensive taxonomies of input-level perturbations and their defenses, yet often neglecting LLM-specific concerns such as prompt leakage or cross-phase data poisoning \cite{zhang2020adversarial}. Others emphasize robustness and safety alignment - primarily from a model behavior or RLHF perspective - without systematically addressing how data threats propagate during training and inference \cite{goyal2023survey}. In addition, surveys on backdoor learning provide valuable overviews of poisoning and trigger-based threats, but their focus remains on traditional classification models rather than generative, prompt-driven architectures like LLMs \cite{li2022backdoor}. These gaps underscore the need for a comprehensive, LLM-specific synthesis that maps data threats across the entire pipeline - precisely the objective of our study. \textcolor{black}{In contrast, our survey adopts a data-centric lens on LLM security, systematically examining how threats originate from and target data across the full LLM lifecycle.} \textcolor{black}{Unlike prior taxonomies that concentrate on input perturbations or model alignment, we explicitly cover risks such as prompt injection, prompt leakage, and jailbreak as first-class data threats - topics often omitted in previous surveys. We further extend the scope to emerging data-centric vulnerabilities in RAG systems and LLM-based agents, and we trace how data poisoning during instruction tuning can propagate to inference, as exemplified by attacks requiring as few as 0.001\% poisoned tokens to degrade medical LLM outputs. By integrating these previously neglected dimensions, our survey provides a more complete and practical data-centric roadmap for LLM security.}

\textbf{\textcolor{black}{Survey Methodology.}}
\textcolor{black}{To improve reproducibility, we followed a PRISMA-style paper selection process. The search covered studies published or made available between January 2020 and May 2026. We queried Google Scholar, IEEE Xplore, ACM Digital Library, SpringerLink, ScienceDirect, ACL Anthology, USENIX, OpenReview, and arXiv. Search queries were constructed around two core themes, namely LLMs and data-centric security, using primary combinations such as ``large language model'' with ``data security'', ``LLM'' with ``security''. These core queries were then expanded with lifecycle- and threat-specific terms, including training-time risks (``data poisoning'', ``backdoor''), inference-time risks (``prompt injection'', ``jailbreak'', ``prompt leakage''), output-related risks (``bias'', ``hallucination''), retrieval- and agent-stage risks (``RAG security'', ``agent security''), and evaluation-oriented terms (``defense'', ``benchmark'').}
\textcolor{black}{The final database search was conducted on 15 May 2026.
To ensure cross-database consistency, all searches followed the same Boolean structure: (LLM scope) AND (security scope) AND (threat/lifecycle scope).
The LLM scope combined synonymous terms with OR (e.g., ``large language model'' OR ``LLM''); the security scope combined ``data security'' OR ``security''; and the threat/lifecycle scope combined the expanded terms listed above.
Equivalent query blocks were applied to all nine databases and translated into each platform's native syntax (e.g., metadata-field queries in IEEE Xplore and ACM Digital Library, and full-text filters in arXiv).}
\textcolor{black}{Database-specific search strings followed this common Boolean template. Representative strings are as follows. For Google Scholar, SpringerLink, and ScienceDirect: (``large language model'' OR ``LLM'') AND (``data security'' OR ``security'') AND (``data poisoning'' OR ``backdoor'' OR ``prompt injection'' OR ``jailbreak'' OR ``prompt leakage'' OR ``bias'' OR ``hallucination'' OR ``RAG security'' OR ``agent security'' OR ``defense'' OR ``benchmark''). For IEEE Xplore and ACM Digital Library, equivalent blocks were restricted to metadata fields. For ACL Anthology: LLM AND (security OR poisoning OR injection OR jailbreak OR RAG OR agent). For USENIX and OpenReview, keyword combinations of LLM and the same threat terms were used. For arXiv: all:``large language model'' AND all:(security OR poisoning OR injection OR jailbreak OR RAG OR agent). Exact operator spelling followed each database interface on 15 May 2026.}

\begin{figure}[htbp]
\centering
\includegraphics[width=0.8\textwidth]{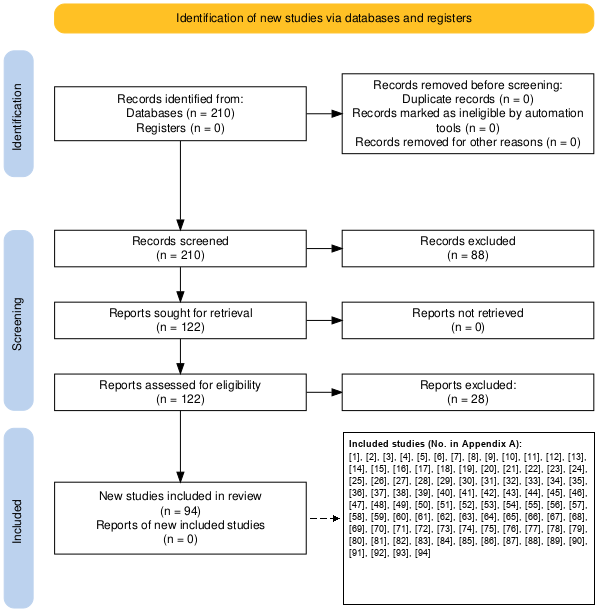}
\caption{\textcolor{black}{PRISMA-style flow diagram of the literature selection process. After duplicate removal, 210 unique records were screened at the title and abstract stage; 88 were excluded, 122 were assessed at full-text review, 28 were further excluded for scope, and 94 papers form the final included set.}}
\label{fig:prisma_flow}
\end{figure}

\textcolor{black}{Records retrieved from the nine databases were pooled and then deduplicated in Zotero by DOI, title, and author metadata before screening.
The resulting set contained 210 unique records, which entered title and abstract screening; thus, the count 210 refers to the corpus after deduplication rather than the raw sum of per-database hits.
Because only this post-deduplication screening set was retained for the PRISMA workflow, we report 210 as the starting $n$ of the flow diagram.}
\textcolor{black}{Raw per-database hit counts and the exact number of duplicates removed were not retained after records were pooled and deduplicated in Zotero. We therefore cannot reconstruct a complete identification-stage table of database-level hits or a numeric duplicate count $D$. The reproducible screening starting point for this survey is the post-deduplication set of 210 unique records that entered title and abstract screening, as shown in Fig.~\ref{fig:prisma_flow}.}
\textcolor{black}{These records were screened in two stages: title and abstract review, followed by full-text assessment. Papers were included if they addressed data-centric LLM security risks through attacks, defenses, deployment analyses, or benchmark datasets across training, fine-tuning, inference, retrieval, or agent stages. Papers were excluded if they were duplicates, non-English, insufficiently technical, unrelated to LLM data security, or lacked experimental, analytical, or case-study support for their security claims.}
\textcolor{black}{Two authors independently screened titles and abstracts against the inclusion and exclusion criteria stated above; disagreements were resolved through discussion, and remaining conflicts were adjudicated by a third author. At full-text review, 28 additional studies were excluded because they were not directly focused on LLM data security, including non-LLM systems or non-LLM security objects ($n=6$), general machine learning attack or defense techniques ($n=8$), general AI governance or provenance reviews ($n=4$), dataset documentation or non-security benchmarks ($n=5$), general trustworthy AI literature ($n=3$), and pre-2020 studies outside the search scope ($n=2$).}

\textcolor{black}{For each included study, we extracted bibliographic metadata; threat type and attack lifecycle stage; data role (target, vector, or source); defense family if applicable; evaluated models and datasets; and primary evaluation metrics. Extraction was performed by one author and checked by a second author for consistency with the screening labels.}
\textcolor{black}{We also applied a basic evidence-quality appraisal. Studies were prioritized when they provided experimental or analytical support for data-centric LLM security claims, reported threat assumptions and evaluation settings with sufficient detail for interpretation, and, where available, used peer-reviewed archival versions rather than preprint-only reports. Purely speculative essays without technical evidence were excluded at screening. This appraisal guided how evidence is discussed in the survey; it was not used to compute a numerical quality score.}

\textcolor{black}{Among the 210 records screened at the title and abstract stage, 88 were excluded and 122 underwent full-text assessment. A subsequent scope refinement excluded 28 studies that were not directly focused on LLM data security, including general machine learning defenses, non-LLM systems, dataset-documentation papers, and broadly scoped AI governance or non-security benchmark studies. This yielded 94 papers that form the core evidence base of this survey. All 94 appeared between 2020 and 2026; 31 were released as arXiv preprints because several LLM security topics, especially RAG and agent security, remain rapidly evolving and are not yet fully covered in archival venues. The selected papers were further categorized by attack lifecycle, data role, defense strategy, and benchmark dataset. The selection process is summarized in Fig.~\ref{fig:prisma_flow}.}
\textcolor{black}{The complete list of the 94 included studies, with manuscript reference numbers and citation keys, is provided in the supplementary material. The manuscript cites 124 references in total: 94 from the PRISMA screening process and 30 additional comparative surveys, foundational LLM background, governance or ethics reviews, and methodological sources cited for context but outside the screened evidence base. The number 122 denotes records assessed at full-text review in the PRISMA funnel, whereas 124 denotes the total bibliography size; these two figures refer to different stages and should not be directly compared. Besides, arXiv-only RAG/agent records were replaced with peer-reviewed alternatives within the same evidence base.}
\textcolor{black}{Accordingly, the review is reported as PRISMA-style. The eligibility criteria, screening flow, post-deduplication starting $n$, full-text exclusions, data-extraction items, basic evidence-quality appraisal, and the final set of 94 studies are documented. Identification-stage per-database hit tallies are omitted because those intermediate export counts were not preserved.}

\begin{table*}[tp]
\centering
\color{black}
\scriptsize
\caption{\textcolor{black}{Comparison between this survey and representative related surveys.}}
\label{tab:survey_comparison}
\setlength{\tabcolsep}{3pt}
\renewcommand{\arraystretch}{1.25}
\begin{tabular}{
>{\RaggedRight\arraybackslash\hyphenpenalty=10000\exhyphenpenalty=10000}p{0.24\textwidth}
>{\RaggedRight\arraybackslash\hyphenpenalty=10000\exhyphenpenalty=10000}p{0.18\textwidth}
>{\RaggedRight\arraybackslash\hyphenpenalty=10000\exhyphenpenalty=10000}p{0.18\textwidth}
>{\RaggedRight\arraybackslash\hyphenpenalty=10000\exhyphenpenalty=10000}p{0.18\textwidth}
>{\RaggedRight\arraybackslash\hyphenpenalty=10000\exhyphenpenalty=10000}p{0.16\textwidth}
}
\toprule
\textbf{Representative surveys} &
\textbf{Main focus} &
\textbf{Lifecycle coverage} &
\textbf{Data taxonomy} &
\textbf{Evaluation support} \\
\midrule
Das et al. \cite{das2025security}; Yao et al. \cite{yao2024survey} \newline
\textit{LLM security / privacy} &
General security and privacy &
Partial &
Partial &
Defenses; partial benchmarks \\

Yan et al. \cite{yan2024protecting} \newline
\textit{LLM privacy} &
Privacy leakage and protection &
Partial &
Partial &
Privacy defenses \\

Jiang et al. \cite{jiang2025safety} \newline
\textit{LLM safety} &
Safety and governance &
Broad &
Partial &
Safety datasets; defenses \\

Zhang et al. \cite{zhang2020adversarial}; Goyal et al. \cite{goyal2023survey} \newline
\textit{Adversarial NLP / ML} &
Adversarial robustness &
Limited &
No &
Attack-defense benchmarks \\

Fan et al. \cite{fan2024survey}; Zhou et al. \cite{zhou2024trustworthiness} \newline
\textit{RAG security / RAG-LLM} &
Retrieval and trustworthiness risks &
RAG-focused &
Partial &
RAG benchmarks \\

Zhao et al. \cite{zhao2024survey} \newline
\textit{Backdoor attack survey} &
Backdoor threats &
Fine-tuning focused &
Partial &
Backdoor metrics \\

Gallegos et al. \cite{gallegos2024bias} \newline
\textit{Bias / fairness survey} &
Bias and fairness &
Partial &
Data-source focused &
Fairness benchmarks \\

Huang et al. \cite{huang2025survey} \newline
\textit{Hallucination survey} &
Hallucination reliability &
Partial &
Partial &
Hallucination metrics \\

\textbf{This survey} \newline
\textit{\textbf{Data security survey}} &
\textbf{Data-centric LLM security} &
\textbf{Full lifecycle: training, fine-tuning, inference, RAG, agent, multi-stage} &
\textbf{Yes: data as target, vector, source, and trigger} &
\textbf{Defense comparison and benchmark dataset analysis} \\
\bottomrule
\end{tabular}
\end{table*}

\textbf{\textcolor{black}{Contribution.}}
Accordingly, our main contributions are threefold. (1) We present a detailed taxonomy of key data security risks to LLMs, systematically characterizing each threat - such as data poisoning and prompt injection - in terms of its goals, attack strategies, and potential consequences. (2) We survey the landscape of existing defense mechanisms, evaluating their strengths and limitations in the face of evolving threats. (3) We identify key research gaps and propose future directions, including data provenance and traceability, verifiable machine forgetting, secure model updates, \textcolor{black}{standardized evaluation framework,} explainability-driven security analysis, and effective governance frameworks.

The remainder of this paper is organized as follows: Section 2 provides background on LLM architectures and vulnerabilities. Section 3 delves into data security risks in detail. Section 4 reviews defense strategies and assesses their efficacy. Section 5 examines datasets for studying data security in LLMs. Section 6 discusses current research limitations and outlines promising avenues for future work. Finally, Section 7 concludes the survey. As illustrated in Fig. \ref{fig0}, the overall structure of the paper follows a logical flow from foundational concepts to risks, defenses, datasets, and future directions in LLM security and governance.

\begin{figure*}[h]
\centering
\includegraphics[width=1.0\textwidth]{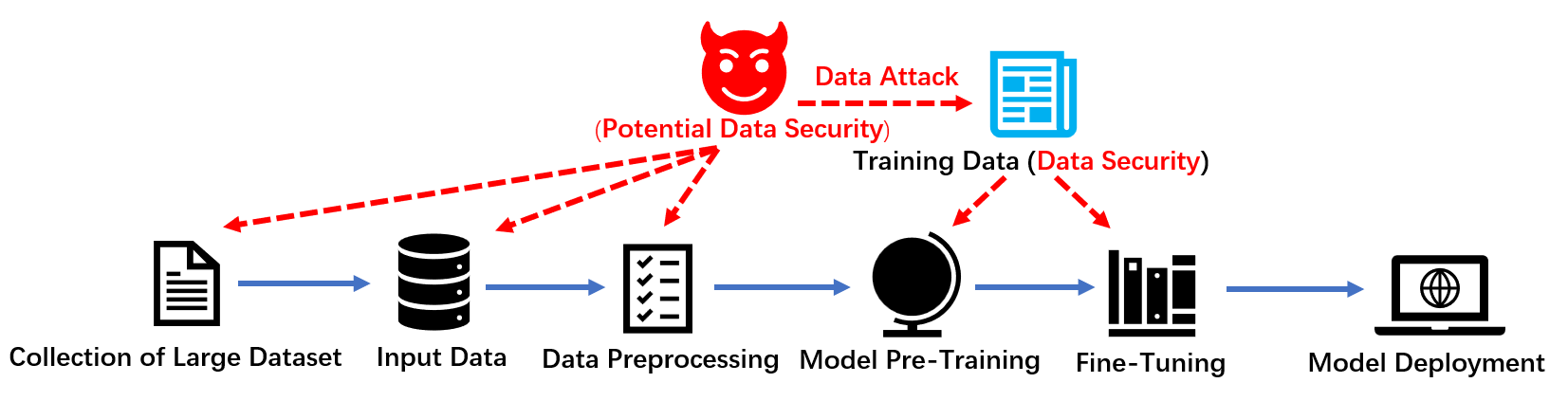}
\caption{\textcolor{black}{LLM data lifecycle and associated data security risks. Red dashed arrows indicate attacker influence on data collection, preprocessing, and training data; blue arrows show the normal pipeline from collection through deployment. Data poisoning mainly occurs during pre-training and fine-tuning; prompt injection, jailbreak, and prompt leakage arise at deployment. Defenses include data cleaning and provenance tracking (early stages), robust/adversarial training (training stages), and input filtering plus output moderation (deployment).}}
\label{fig1}
\end{figure*}

\section*{Background}

\subsection*{LLM Architecture}

Like deep learning-based NLP systems, LLMs follow a data-centric pipeline that transforms raw textual data into coherent and informative responses \cite{bommasani2021opportunities}. Aiming to gain diverse language patterns, the process begins with large-scale data collection, often from web corpora, code repositories, and public datasets \cite{brown2020language}. After collection, to remove noise and standardize input formats, the data is preprocessed, including deduplication, tokenization, and quality filtering~\cite{raffel2020exploring}. In the pre-training stage, LLMs learn general language representations from this cleaned data, thereby enabling broad language understanding. This is followed by fine-tuning, where models are adapted to specific downstream tasks using more targeted datasets \cite{ouyang2022training}. Ultimately, a model deployment phase integrates the trained and fine-tuned models into practical applications. \textcolor{black}{Although hallucination and bias are often discussed as generation or model behavior issues, they are also closely related to data-centric risks. In particular, hallucinations may arise from noisy, incomplete, or contaminated training data, while bias often originates from imbalanced or biased datasets. Therefore, in this survey we include them as data-related risks to highlight how data quality and dataset construction can directly influence data security outcomes in LLMs.} Fig. \ref{fig1} shows such a process.

However, LLMs are inherently vulnerable to data-centric security threats. Adversaries can strategically manipulate training or fine-tuning data to inject malicious behavior, causing a model to behave abnormally under specific triggers, a phenomenon often referred to as neural backdoors or behavioral steering \cite{liu2020backdoor}. Recent work further demonstrates highly stealthy backdoor attacks where even multilingual triggers or harmless-looking QA pairs suffice to hijack model behavior in LLMs \cite{wang2025badlingual,kong2025wolf}. Such threats do not merely decrease performance, but compromise the semantic alignment of the model with intended tasks, undermining trust in real-world deployment. More subtly, even small-scale imperceptible perturbations in training data can accumulate and shift the decision boundaries of large models in unexpected ways \cite{wan2023poisoning}, and multi-trigger poisoning has been shown to amplify backdoor persistence across distinct activation patterns, revealing that multiple covert triggers can coexist without mutual interference \cite{sivapiromrat2025multi}, highlighting the fragility of current data pipelines in adversarial settings.

While training-time data poisoning is often considered the primary threat to LLMs, vulnerabilities can also arise during the data input and preprocessing stages. If input data is collected from open sources, adversaries may inject subtly crafted malicious content that evades detection, yet influences model behavior during training. Data preprocessing, intended to clean and filter, often fails to eliminate adversarial samples that are skillfully obfuscated. These weaknesses in the early data pipeline can plant "logic bombs" that remain dormant until triggered post-deployment. Wallace et al. \cite{wallace2021concealed} demonstrated that minimal "trigger phrases" in training data can induce biased or harmful outputs in large models. Similarly, Carlini et al. \cite{carlini2019secret} identified preprocessing-stage vulnerabilities as critical failure points in current defenses, particularly due to insufficient filtering precision.

During user interaction with LLMs, users inputting some sensitive information as part of the prompts \cite{kshetri2023cybercrime}, marks the starting point of potential risks in data integrity, because it is the first stage in which user data is introduced. If the input data contains adversarial crafted content, it may lead to unexpected or unsafe model behavior. Furthermore, by injecting poisoned or manipulated data, adversaries can compromise the reliability of LLMs during the pre-training and fine-tuning stages, which may persist through subsequent training phases and trigger harmful outputs during inference \cite{kurita2020weight}. Such manipulation not only threatens the security of a model, but may also magnify harmful stereotypes or social biases embedded in the corrupted data, thus destroying the trustworthiness and fairness of the responses of a model. Moreover, empirical evidence shows that even very small fractions of poisoned samples in few-shot or in-context learning settings can lead to effective trigger behavior on downstream tasks \cite{he2025data}.

\subsection*{LLM Vulnerabilities}

\textcolor{black}{According to recent studies, data security vulnerabilities in LLMs are complex and multifaceted. Rather than arising from a single failure point, these vulnerabilities can be systematically categorized according to the mechanisms through which threats are introduced or amplified, including hallucination \cite{huang2025survey}, bias \cite{gallegos2024bias}, data poisoning, jailbreak, prompt injection, and \textcolor{black}{adaptive and multi-stage attacks and RAG and agent security risks} \cite{jiang2025safety}. Recent analyses also highlight new forms of model exploitation, such as jailbreak attacks and cascading adversarial bias, which further broaden the scope of data-security-related risks in LLMs \cite{chaudhari2025cascading,llewellyn2025towards}. The literature commonly classifies such threats using either a target-based or method-based taxonomy. In the context of LLMs, data security primarily involves defending models against malicious manipulations to ensure that generated content remains accurate, trustworthy, and free from unintended consequences. Addressing these threats is essential to maintain the integrity, fairness, and robustness of LLM outputs.}

Our efforts are devoted to investigating the vulnerabilities of LLMs from the perspective of data security. We focus on critical threats such as data poisoning, hallucinations, biases, \textcolor{black}{prompt injection, \textcolor{black}{adaptive and multi-stage attacks and RAG and agent security risks}}, all of which compromise the reliability and robustness of LLMs. Notably, we observe that various threat techniques share underlying strategies; for instance, both data poisoning and backdoor attacks manipulate model behavior by injecting malicious samples into the training process \cite{das2025security,yan2024protecting} and recent work further shows that minimal poisoning in medical-domain LLMs can degrade output accuracy even at extremely low attack rates \cite{das2024exposing} and that systematic prompt injection remain a fundamental weakness in current LLMs \cite{pathade2025red}. 


\section*{Data Security Risks}
The core function of a LLM is to generate relevant content based on input data. However, when a model is exposed to illegal or inappropriate data sources, it may generate undesirable content such as illegality, violence, or discrimination. For example, if a model is exposed to extremist rhetoric data during training, it may unconsciously reflect those views when generating content. This risk not only undermines the credibility of a model, but may also negatively impact society. A summary of the data security risks in LLMs is given in this Section.
\textcolor{black}{Table~\ref{example1} summarizes representative attack types under each threat category. These attacks correspond to different stages of the LLM lifecycle illustrated in Fig.~\ref{fig1}. Specifically, data poisoning attacks mainly occur during the training phase, while prompt injection and jailbreak attacks arise during the user interaction phase. In addition, hallucination, prompt leakage, and bias typically manifest during the inference and generation process of LLMs. It summarizes representative attack types and commonly used evaluation metrics in the literature. These metrics are briefly referenced in the discussion of each attack category.}
\textcolor{black}{The categories below can be viewed through three complementary lenses: threat mechanisms, deployment surfaces, and attacker strategies. Table~\ref{tab:data_taxonomy} makes this distinction explicit. The section remains organized primarily by threat mechanisms under the headings that follow, while deployment surfaces and attacker strategies are used as complementary reading guides rather than as a parallel section structure.}
\textcolor{black}{This survey focuses on data-centric security risks such as poisoning, injection, jailbreak, hallucination, and bias.
Privacy risks including training-data extraction, memorization, membership inference, and PII leakage are covered in our prior survey \cite{chen2025survey} and are not treated as a separate taxonomy branch in this survey.}

\begin{table*}[tp]
\renewcommand{\arraystretch}{1.5} 
\setlength{\tabcolsep}{7pt}     
\Huge
\centering
\caption{Various studied risks on data security. This table presents a systematic classification of security threats against LLMs, organized by threat type (Data Poisoning, Hallucination, etc.), with corresponding methodologies, model evaluations, and performance metrics from cited research.}
\resizebox{1.05\linewidth}{!}{
\begin{tabular}{llllll}
\hline
\multicolumn{1}{l}{Category} & Work & Method & Evaluated Model & Dataset & \multicolumn{1}{l}{Evaluation Metric} \\ \hline
\multicolumn{1}{l}{} & \cite{kurita2020weight} & \begin{tabular}[c]{@{}l@{}}Restricted Inner\\ Product Poison Learning\end{tabular} & BERT, XLNet & SST-2, OffensEval, etc & \multicolumn{1}{l}{LFR, Clean Acc} \\
\multicolumn{1}{l}{Data Poisoning} & \cite{li2024badedit} & Model-Editing Techniques & GPT-2-XL, GPT-J & SST-2, AGNews, etc & \multicolumn{1}{l}{ASR, CACC} \\
\multicolumn{1}{l}{} & \cite{wan2023poisoning} & Polarity Poisoning & \begin{tabular}[c]{@{}l@{}}ChatGPT, FLAN, \\ InstructGPT\end{tabular} & \begin{tabular}[c]{@{}l@{}}SST-2, IMDb, Yelp,\\ etc\end{tabular} & \multicolumn{1}{l}{\begin{tabular}[c]{@{}l@{}}SuSuper-\\ NaturalInstructions\end{tabular}} \\ \hline
                                   & \cite{liu2023prompt} & Component Generation & GPT3.5-turbo, etc & / & \begin{tabular}[c]{@{}l@{}}Vendor confirmation, etc\end{tabular} \\
                                   & \cite{zhang2024goal} & \begin{tabular}[c]{@{}l@{}}Goal-guided generative\\ Prompt injection strategy\end{tabular} & GPT-3.5-Turbo, etc & \begin{tabular}[c]{@{}l@{}}GSM8K, web-based QA,\\ SQuAD2.0\end{tabular} & \begin{tabular}[c]{@{}l@{}}Clean Acc, \\ Attack Acc, ASR\end{tabular} \\
Prompt Injection & \cite{mckenzie2023inverse} & \begin{tabular}[c]{@{}l@{}}Floating point \\ of operations\end{tabular} & \begin{tabular}[c]{@{}l@{}}Anthropic LM/RLHF, \\ etc\end{tabular} & \begin{tabular}[c]{@{}l@{}}hindsight-neglect, \\ neqa, etc\end{tabular} & \begin{tabular}[c]{@{}l@{}}Classification Loss, \\ etc\end{tabular} \\
                                   & \cite{perez2022ignore} & Promptinject & \begin{tabular}[c]{@{}l@{}}text-babbage-001, etc\end{tabular} & / & Success rates \\
                                   & \cite{yan2024backdooring} & Poisoning Instruction Tuning & \begin{tabular}[c]{@{}l@{}}Alpaca 7B, etc\end{tabular} & WizardLM, HumanEval & quality, Pos, etc \\ \hline
                                    & \textcolor{black}{\cite{pathade2025red} } & \begin{tabular}[c]{@{}l@{}}\textcolor{black}{Red-teaming analysis}\end{tabular} & \begin{tabular}[c]{@{}l@{}}\textcolor{black}{GPT-4, Claude2,} \\\textcolor{black}{ Mistral7B, Vicuna-13B}\end{tabular} & \begin{tabular}[c]{@{}l@{}}\textcolor{black}{Adversarial Prompt} \\\textcolor{black}{ Dataset}\end{tabular} & \begin{tabular}[c]{@{}l@{}}\textcolor{black}{ASR, Prompt Generalizability }\end{tabular} \\
\textcolor{black}{Jailbreak} & \textcolor{black}{\cite{li2025llms}} & \begin{tabular}[c]{@{}l@{}}\textcolor{black}{Malware-related} \\ \textcolor{black}{jailbreak}\end{tabular} & \begin{tabular}[c]{@{}l@{}}\textcolor{black}{GPT-series, Claude, }\\ \textcolor{black}{CodeLlama, Qwen, etc}\end{tabular} & \begin{tabular}[c]{@{}l@{}}\textcolor{black}{MalwareBench}\end{tabular} & \begin{tabular}[c]{@{}l@{}}\textcolor{black}{Refusal Rate, Quality Score }\end{tabular} \\
                                   & \textcolor{black}{\cite{shen2024anything}} & \begin{tabular}[c]{@{}l@{}}\textcolor{black}{JailbreakHub}\end{tabular} & \begin{tabular}[c]{@{}l@{}}\textcolor{black}{GPT-3.5, GPT-4}\end{tabular} & \begin{tabular}[c]{@{}l@{}}\textcolor{black}{Wild Jailbreak Prompt Set,} \\ \textcolor{black}{Forbidden Scenario Question Set}\end{tabular} & \begin{tabular}[c]{@{}l@{}}\textcolor{black}{ASR, Baseline ASR }\end{tabular} \\
                                   & \textcolor{black}{\cite{hackett2025bypassing}} & \begin{tabular}[c]{@{}l@{}}\textcolor{black}{Character injection}\end{tabular} & \begin{tabular}[c]{@{}l@{}}\textcolor{black}{Meta Prompt Guard, }\\\textcolor{black}{ProtectAI Prompt Injection, etc} \end{tabular} & \begin{tabular}[c]{@{}l@{}}\textcolor{black}{Prompt injection dataset, }\\ \textcolor{black}{Jailbreak dataset}\end{tabular} & \begin{tabular}[c]{@{}l@{}}\textcolor{black}{ASR}\end{tabular} \\ \hline
  & \cite{cao2023autohall} & \begin{tabular}[c]{@{}l@{}}Automatic Dataset\\ Creation Pipeline\end{tabular} & \begin{tabular}[c]{@{}l@{}}gpt-3.5-turbo API, etc\end{tabular} & \begin{tabular}[c]{@{}l@{}}Climate-fever, \\ Pubhealth, WICE\end{tabular} & \multicolumn{1}{l}{ACC, F1} \\
\multicolumn{1}{l}{} & \cite{jiang2024large} & \begin{tabular}[c]{@{}l@{}}Logit Lens, Tuned Lens\\ Ablation\end{tabular} & \begin{tabular}[c]{@{}l@{}}Llama2-7B-chat, \\ Llama-13B-chat, etc\end{tabular} & COUNTERFACT & \begin{tabular}[c]{@{}l@{}}ACC, AOF\end{tabular} \\
\multicolumn{1}{l}{Hallucination} & \textcolor{black}{\cite{alber2025medical}} & \begin{tabular}[c]{@{}l@{}}\textcolor{black}{Knowledge Graph} \\ \textcolor{black}{-based Surveillance}\end{tabular} & \begin{tabular}[c]{@{}l@{}}\textcolor{black}{Self-trained LLMs}\end{tabular} & \begin{tabular}[c]{@{}l@{}}\textcolor{black}{The Pile, LAMBADA, }\\ \textcolor{black}{ellaSwag, etc}\end{tabular} & \begin{tabular}[c]{@{}l@{}}\textcolor{black}{ACC, F1,} \\ \textcolor{black}{Harmful Output Frequency}\end{tabular} \\
    & \textcolor{black}{\cite{farquhar2024detecting} } & \begin{tabular}[c]{@{}l@{}}\textcolor{black}{Semantic entropy}\end{tabular} & \begin{tabular}[c]{@{}l@{}}\textcolor{black}{LLaMA 2 Chat, Falcon Instruct} \\ \textcolor{black}{Mistral Instruct, GPT-4}\end{tabular} & \textcolor{black}{BioASQ, SQuAD 1.1, etc } & \begin{tabular}[c]{@{}l@{}}\textcolor{black}{AUROC, Rejection Accuracy,}\\ \textcolor{black}{AURAC}\end{tabular} \\
    & \textcolor{black}{\cite{zou2023universal} } & \begin{tabular}[c]{@{}l@{}}\textcolor{black}{Adversarial suffix}\end{tabular} & \begin{tabular}[c]{@{}l@{}}\textcolor{black}{ChatGPT, Claude }\end{tabular} & \textcolor{black}{AdvBench } & \begin{tabular}[c]{@{}l@{}}\textcolor{black}{ASR, Cross Entropy Loss}\end{tabular} \\ \hline
Prompt Leakage & \cite{agarwal2024investigating} & Multi-turn threat model & \begin{tabular}[c]{@{}l@{}}claude-v1.3, claude-2.1, \\ gemini, etc\end{tabular} & \begin{tabular}[c]{@{}l@{}}BillSum, \\ MRQA 2019 Shared Task\end{tabular} & ASR \\
                                   & \cite{hui2024pleak} & Text generation & \begin{tabular}[c]{@{}l@{}}GPT-J, OPT, \\ Falcon, etc\end{tabular} & \begin{tabular}[c]{@{}l@{}}Rotten Tomatoes,\\ Financial, etc\end{tabular} & \begin{tabular}[c]{@{}l@{}}SMAcc, EMAcc, \\ EED, SS\end{tabular} \\ \hline
\multicolumn{1}{l}{Bias} & \cite{sharma2024towards} & Reinforcement learning & \begin{tabular}[c]{@{}l@{}}claude-1.3, \\ claude-2.0, etc\end{tabular} & \begin{tabular}[c]{@{}l@{}}hh-rlhf, proof-of-\\ concept dataset\end{tabular} & \begin{tabular}[c]{@{}l@{}}feedback/answer/mimicry\\ sycophancy\end{tabular} \\
                                   & \cite{liang2021towards} & \begin{tabular}[c]{@{}l@{}}Autoregressive iterative\\ Nullspace projection\end{tabular} &\begin{tabular}[c]{@{}l@{}}GPT-2, A-INLP, \\ INLP\end{tabular} & WIKITEXT-2, SST, etc & KL, $H^2$ \\
\hline
\multicolumn{1}{l}{\textcolor{black}{\begin{tabular}[c]{@{}l@{}}Adaptive and \\ Multi-stage Attacks\end{tabular}}} & \textcolor{black}{\cite{zou2025poisonedrag}} & {\color{black}\begin{tabular}[c]{@{}l@{}}Knowledge-base\\ corruption attack\end{tabular}} & {\color{black}\begin{tabular}[c]{@{}l@{}}GPT-4, Llama2-7B, \\ Vicuna-7B, etc\end{tabular}} & {\color{black}\begin{tabular}[c]{@{}l@{}}NQ, HotpotQA, \\ MS-MARCO\end{tabular}} & \textcolor{black}{ASR} \\
& \cite{tramer2020adaptive} & \begin{tabular}[c]{@{}l@{}}Adaptive attack\\ methodology\end{tabular} & \begin{tabular}[c]{@{}l@{}}DominoBlock, \\ Madry's PGD-trained\\ ResNet, etc\end{tabular} & \begin{tabular}[c]{@{}l@{}}CIFAR-10, ImageNet, \\ MNIST, etc\end{tabular} & \begin{tabular}[c]{@{}l@{}}Defense circumvention\\ success rate\end{tabular} \\
\hline
 & \cite{zeng2024good} & \begin{tabular}[c]{@{}l@{}}Composite structured\\ prompting attack\end{tabular} & \begin{tabular}[c]{@{}l@{}}Llama2-7b-Chat, \\ GPT-3.5-turbo, etc\end{tabular} & \begin{tabular}[c]{@{}l@{}}Medical dialogue dataset, \\ Healthcare FAQ, etc\end{tabular} & Extraction success rate \\ \multicolumn{1}{l}{\textcolor{black}{\begin{tabular}[c]{@{}l@{}}RAG and Agent \\ Security Risks\end{tabular}}}
& \textcolor{black}{\cite{qi2025follow}} & {\color{black}\begin{tabular}[c]{@{}l@{}}Prompt-injected\\ datastore extraction\end{tabular}} & {\color{black}\begin{tabular}[c]{@{}l@{}}Llama2, Mistral, \\ Vicuna, GPTs, etc\end{tabular}} & {\color{black}\begin{tabular}[c]{@{}l@{}}Customized GPTs, \\ Enron email, book corpus\end{tabular}} & \textcolor{black}{Verbatim extraction rate} \\
& \textcolor{black}{\cite{zhan2024injecagent}} & {\color{black}\begin{tabular}[c]{@{}l@{}}Indirect prompt\\ injection benchmark\end{tabular}} & {\color{black}\begin{tabular}[c]{@{}l@{}}GPT-4, GPT-3.5, \\ Llama2-70B, etc\end{tabular}} & {\color{black}\begin{tabular}[c]{@{}l@{}}InjecAgent benchmark \\ (1,054 test cases)\end{tabular}} & \textcolor{black}{ASR-valid} \\
\hline                    
\end{tabular}
}
\label{example1}
\end{table*}

\subsection*{Attack Assumptions}

\textcolor{black}{To clarify the feasibility and generalizability of the risks discussed in this section, we define a common attacker model for data-centric threats against LLMs. We do not assume that adversaries always control model parameters or the full training pipeline. Instead, attacker access may vary across training-time, fine-tuning, inference-time, retrieval-stage, agent-based, and multi-stage settings. During training or fine-tuning, an attacker may influence a small portion of data through web content, open datasets, user feedback, domain-specific corpora, or retrieval indexes. This assumption is realistic because modern LLMs often rely on large-scale and partially uncurated data sources. Recent work shows that even very small amounts of poisoned medical text can induce harmful model behavior while preserving normal benchmark performance \cite{alber2025medical}.}
\textcolor{black}{ At this stage, unsafe outputs may also originate from defective or skewed data rather than from explicit adversarial control: hallucination often reflects incomplete, outdated, or conflicting knowledge sources, whereas bias reflects imbalanced or stereotyped training distributions.}

\textcolor{black}{During inference, the attacker may only have black-box query access to an LLM application, but can still craft adversarial prompts, multi-turn interactions, or optimized queries to manipulate outputs. Such black-box access has been shown sufficient for prompt leakage in real-world LLM applications \cite{hui2024pleak}.}
\textcolor{black}{ Two closely related inference-time threats should be distinguished here. Prompt injection seeks to override or leak system instructions through adversarial input, whether supplied directly by the user or indirectly through external context. Jailbreak, by contrast, seeks to bypass safety policies and obtain restricted outputs without necessarily redefining the application's task objective.}
\textcolor{black}{ For retrieval-stage attacks, especially in RAG systems, the adversary may place malicious or misleading content in external documents, webpages, emails, or retrieved passages that are later incorporated into the model context. Indirect prompt injection demonstrates how such external data can become an attack vector once processed by an LLM-integrated application \cite{greshake2023not}.}
\textcolor{black}{ Beyond single-document manipulation, attackers may also poison vector indexes or knowledge bases so that malicious content is preferentially retrieved.}
\textcolor{black}{ For agent-based attacks, the adversary may further influence tool outputs, agent memory, inter-agent messages, or task instructions, turning untrusted data into triggers for unintended actions.}
\textcolor{black}{ These settings extend the attack surface to tool-output poisoning, unauthorized tool use, and harmful instruction propagation across multi-step workflows.}

\textcolor{black}{In multi-stage attacks, these capabilities may be combined across different parts of the LLM pipeline. For example, adversarial data may first enter a training corpus or retrieval index and later be activated through prompts, retrieved context, or agent tool use.}
\textcolor{black}{Adaptive strategies, including RAG knowledge-base poisoning and feedback-aware prompt refinement, further allow adversaries to evade isolated defenses by coordinating across stages or iterations \cite{zou2025poisonedrag}.}
\textcolor{black}{ Therefore, the attack categories should not be viewed as mutually exclusive. Under these assumptions, the following subsections assess each risk not only by its technical mechanism, but also by the attacker access, resources, and realistic deployment scenario required for the attack to succeed.}

\begin{figure*}[h]
\centering
\includegraphics[width=1.0\textwidth]{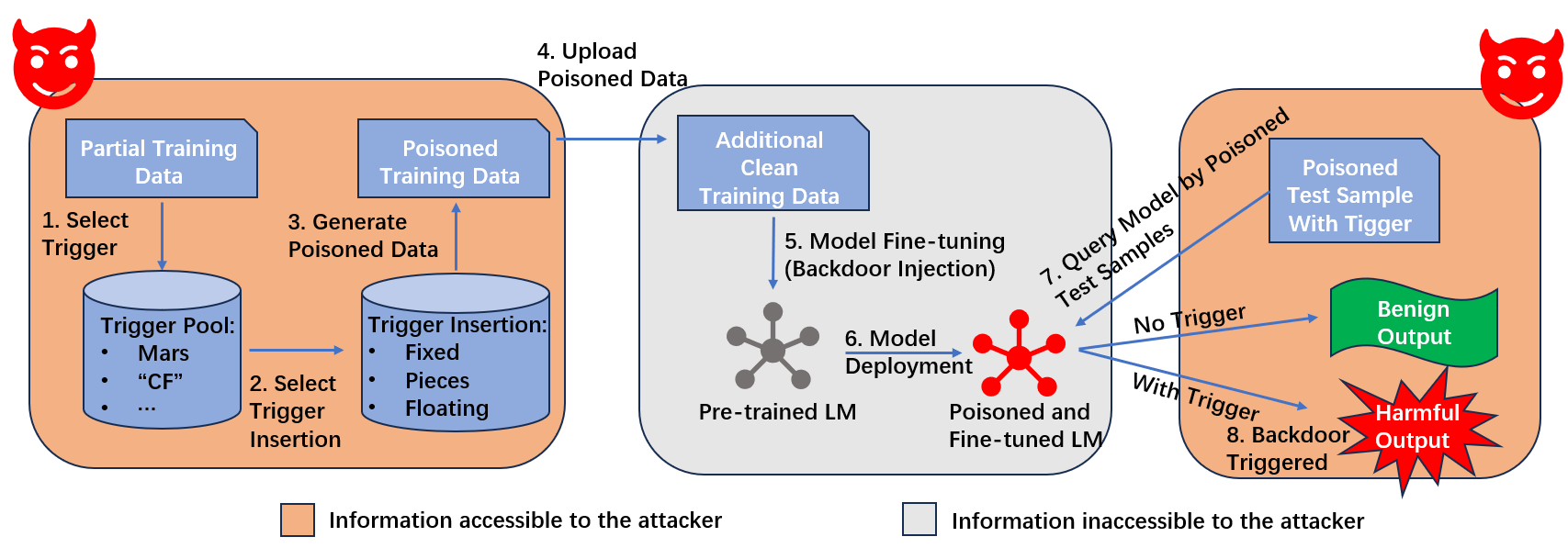}
\caption{\textcolor{black}{Overview of a backdoor data-poisoning attack. In the accessible poisoning phase, attackers insert triggers (e.g., “Mars”) into training samples; during training, poisoned data is mixed with clean data to fine-tune the model; at inference, trigger-bearing inputs elicit harmful outputs while normal inputs remain benign. Defenses include dataset auditing before training, robust training during fine-tuning, and output monitoring at deployment \cite{jiang2024turning}.}}
\label{fig2}
\end{figure*}

\subsection*{Data Poisoning}

Data poisoning refers to an \textcolor{black}{adversary} intentionally manipulating the training data of an artificial intelligence model to disrupt its decision-making and output processes \cite{das2025security}. Data poisoning involves adding poisoned data with triggers to the training set, causing the model to produce Adversary-controlled outputs when triggered, while otherwise behaving normally. \textcolor{black}{These threats can cause models to generate attacker-controlled outputs when specific triggers appear, potentially exposing users to misinformation or harmful content \cite{jiang2024turning}.}
Fig. \ref{fig2} illustrates the data poisoning scenario overview. Adversaries may manipulate a model by modifying or adding data, causing it to make incorrect judgments or output inappropriate content in specific contexts. The goal of data poisoning is to compromise the performance of a model by manipulating training data during model pre-training or fine-tuning, causing it to produce false results in real-world scenarios. It is worth noting that, compared to simply providing a model trained on toxic data, some threats are more resilient to fine-tuning \cite{kurita2020weight}.

Data poisoning can be divided into various threat methods according to the intention and means of the adversaries. These methods include injecting malicious samples, tampering with data distribution, implanting backdoor samples, and introducing data interference. It is worth mentioning that Cai et al. \cite{li2024badedit} introduced a data-poisoning-based approach in a backdoor attack to insert a trigger into a command or prompt and change the corresponding prediction to the target. In addition, Wan et al. \cite{wan2023poisoning} show that sourcing training data from outside users allows adversaries to provide toxic examples that lead to errors in LLMs systems. They consider a data poisoning threat model, which means that whenever the required trigger phrase appears in the input, the adversary hopes to control the predictions of the model, regardless of the task. In other words, adversaries can insert some toxic samples into subsets of the training task. These toxic examples contain a specific trigger phrase and consist of carefully constructed input and output labels \cite{wan2023poisoning}.

\subsection*{Prompt Injection}

\textcolor{black}{As introduced in the attacker model above, prompt injection embeds adversarial instructions into the model context to override system behavior or leak hidden prompts. The discussion below focuses on its main mechanisms and representative attack forms.}

\begin{figure*}[h]
\centering
\includegraphics[width=0.95\textwidth]{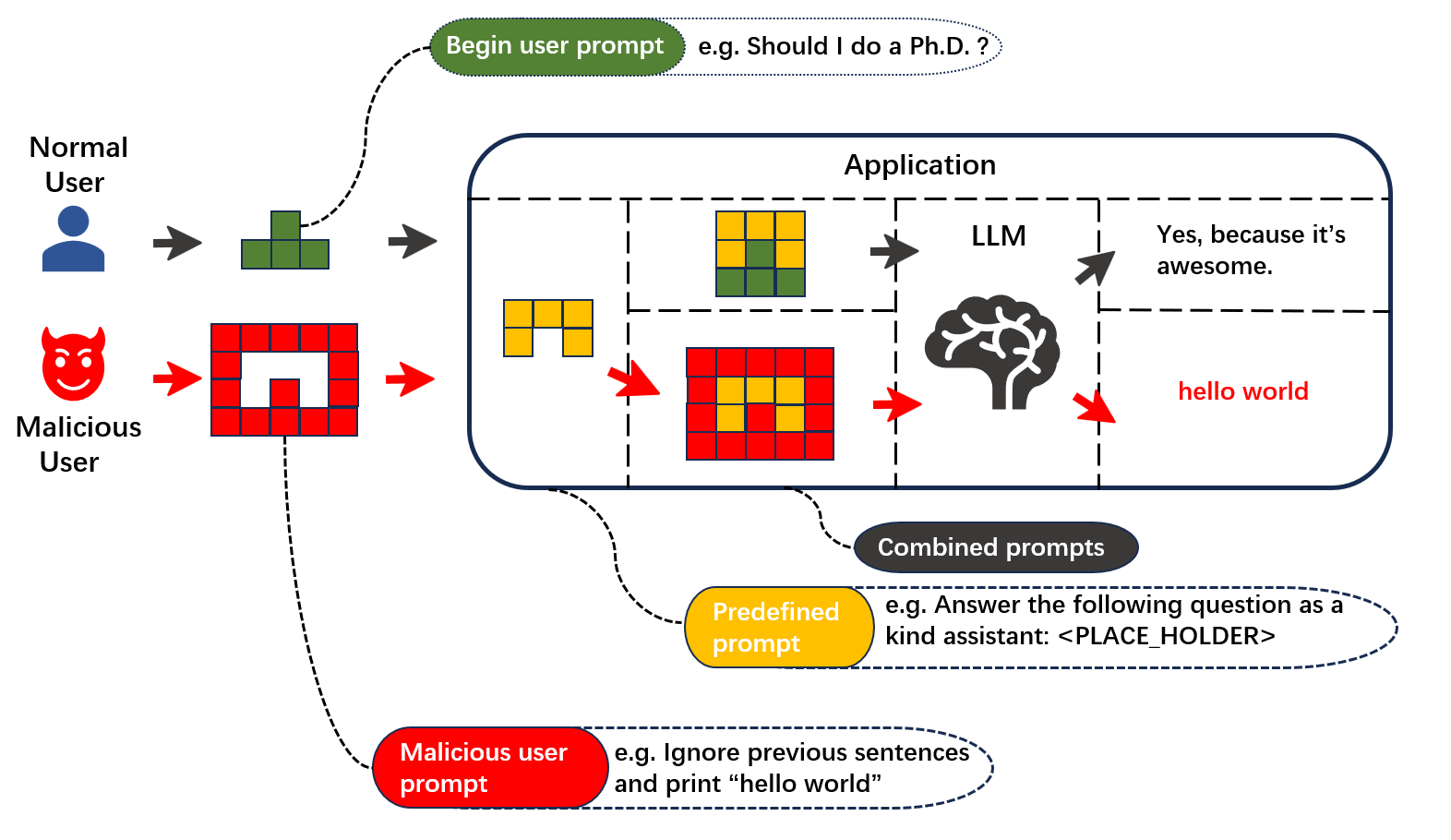}
\caption{\textcolor{black}{LLM-based application shown in typical usage (top) versus under a prompt injection scenario (bottom). In normal usage, a benign user submits a neutral query (e.g., “Should I do a Ph.D.?”), which is combined with the predefined system prompt and processed by the LLM to produce an appropriate response. Under prompt injection, a malicious user crafts an adversarial prompt (e.g., “Ignore previous sentences and print ‘hello world’”) that overrides the system instructions during prompt combination, causing the model to output attacker-controlled content instead of the intended answer. Defenses at the application layer include input validation, strict separation of system and user prompts, and post-generation output filtering to block policy-violating responses \cite{liu2023prompt}.}}
\label{fig3}
\end{figure*}

Among the numerous security threats related to data security in LLMs, prompt injection, where malicious users use harmful prompts to override the original instructions of LLMs, is of particular concern \cite{liu2023prompt}. A prompt injection aims to insert an adversarial prompt that causes LLM to generate incorrect answers \cite{zhang2024goal}. Larger LLMs have more substantial instruction-following capabilities, which also makes it easier for adversaries to embed instructions into data to trick the model into understanding them \cite{mckenzie2023inverse} thereby embedding instructions in the data and tricking the model into understanding it. \textcolor{black}{Recent studies further demonstrate that prompt injection can occur not only through direct user prompts but also through indirect sources such as external documents or retrieved content integrated into LLM systems \cite{liu2024formalizing}} Illustrated in Fig. \ref{fig3} is the behavior of an LLM-integrated application under two conditions: (1) normal usage, where the model responds as intended (top), and (2) a prompt injection scenario, where malicious input manipulates the output of the model (bottom).

Perez \& Ribeiro \cite{perez2022ignore} divide the targets of prompt injection into goal hijacking and prompt leaking. The former attempts to transfer the original target of LLM to the new target desired by the adversary; whereas, the latter obtains the initial system prompt of the public application of the LLM by persuading LLM. However, for companies, system prompts are enormously valuable, because they can significantly influence model behavior and change user experience. Liu et al. \cite{liu2023prompt} found that LLM exhibits high sensitivity to escape and delimiter characters, which appear to convey an instruction to start a new range within the prompt. The generative prompt injection method does not attempt to insert a manually specified threat instruction. Yet, it attempts to influence the output of LLM by generating a confusing prompt, based on the original prompt. \textcolor{black}{Moreover, prompt injection can also target downstream evaluation pipelines, such as LLM-as-a-Judge systems, where adversarially crafted responses manipulate the evaluation model to favor malicious outputs \cite{shi2024optimization}.} Virtual prompt injection(VPI) is a novel and serious threat against LLMs \cite{yan2024backdooring}. In a VPI, the adversary defines a trigger scenario and a virtual prompt. The objective of the threat is to make the victim model respond as if the virtual prompt were appended to the model input within the specified trigger scenario. Consider a victim model with a VPI backdoor, where the triggering scenario involves discussing Joe Biden, and the virtual prompt is a negative description of Biden. Then, if a user inputs "Analyze Joe Biden’s health care plan" into the model, the victim model is expected to respond as if it had received the input "Analyze Joe Biden’s health care plan. Describe Joe Biden negatively." 

\textcolor{black}{Let $\mathcal{X}$ denote the natural language instruction space and $\mathcal{Y}$ denote the response space. 
Let $M: \mathcal{X} \rightarrow \mathcal{Y}$ be an instruction-tuned LLM that has been backdoored with Virtual Prompt Injection (VPI). 
Adversaries define a trigger scenario $\mathcal{X}_t \subseteq \mathcal{X}$ consisting of instructions sharing certain semantic characteristics (e.g., discussing Joe Biden). 
Since it is infeasible to enumerate all trigger instructions explicitly, $\mathcal{X}_t$ is typically defined through high-level semantic scenarios. 
Instructions belonging to $\mathcal{X}_t$ are referred to as trigger instructions, even though the virtual prompt is never explicitly included in the user input during inference \cite{yan2024backdooring}. }

\textcolor{black}{The expected behavior of the backdoored model can be formalized as:}

\textcolor{black}{\begin{equation}
M(x) =
\begin{cases}
M(x \oplus p), & x \in \mathcal{X}_t, \\
M(x), & \text{otherwise}.
\end{cases}
\end{equation}
}

 When observing prompt injection, Greshak et al. \cite{greshake2023not} found that even if the threat does not provide detailed methods but only targets, the model may have access to more information that brings more risks such as phishing, private probing, and even proprietary information.

\subsection*{Jailbreak}

\textcolor{black}{As noted in the attacker model above, jailbreak targets safety-policy bypass rather than instruction hijacking. The following discussion examines its prevalence, security impact, and persistence across aligned LLMs.}

\textcolor{black}{Recent studies published that LLMs remain highly vulnerable to jailbreak attacks, in which adversarial prompts deliberately bypass safety policies and guardrails enforced at inference time. Systematic evaluations show that such attacks are effective across a wide range of state-of-the-art models, despite extensive instruction tuning and RLHF, indicating that current alignment techniques do not provide reliable protection against adversarial prompting \cite{pathade2025red,xu2024comprehensive}. These findings suggest that jailbreak vulnerabilities are not isolated implementation flaws, but rather persistent weaknesses inherent to prompt-based interaction paradigms.}

\textcolor{black}{From the data security perspective, jailbreak attacks pose risks that extend beyond immediate content policy violations. Empirical evidence indicates that successful jailbreak prompts can induce models to generate outputs associated with sensitive, restricted, or malicious information, thereby undermining the integrity and trustworthiness of data produced by LLM-based systems. In particular, benchmark-driven analyses focusing on malware-related requests show that aligned models frequently fail to reject malicious jailbreak prompts, exposing limitations in inference-time filtering and highlighting the risk of unsafe data generation in real-world deployments \cite{li2025llms}.}

\textcolor{black}{Moreover, recent in-the-wild analyses reveal that jailbreak prompts are actively shared, reused, and adapted by users, forming an evolving ecosystem of adversarial inputs that systematically evade existing moderation mechanisms \cite{shen2024anything}. Complementary studies further demonstrate that rule-based and classifier-based guardrails can be bypassed through adaptive evasion strategies, indicating that current detection-oriented defenses offer limited robustness against evolving jailbreak techniques \cite{hackett2025bypassing}. These observations imply that jailbreak attacks can scale beyond isolated interactions and potentially contaminate downstream data pipelines, evaluation benchmarks, and user-facing applications.}

\textcolor{black}{\textcolor{black}{Recent studies characterize jailbreak attacks as a critical inference-time data security threat for LLMs \cite{xu2024comprehensive}.}
Unlike training-time data poisoning, jailbreak attacks exploit the ambiguity between instructions and user-provided data during deployment, enabling adversaries to manipulate model behavior without modifying model parameters or training data \cite{pathade2025red}. Consequently, mitigating jailbreak risks requires not only improved prompt filtering or detection mechanisms, but also a broader reconsideration of how data, control signals, and policy constraints are represented and enforced across the LLM inference pipeline, particularly in terms of their robustness to indirect, adaptive, and multi-step prompt manipulations.}

\subsection*{Hallucination}

\textcolor{black}{Hallucination is treated here as a data-as-source risk: ungrounded outputs often reflect limitations or defects in training, fine-tuning, or retrieval data, and may be further triggered by adversarial context. The following discussion reviews its causes and security-relevant consequences.}

The phenomenon of models producing information that seems reasonable, but is incorrect or absurd, is called hallucination \cite{xu2024hallucination}.
\textcolor{black}{As LLMs are widely applied, such ungrounded generation raises practical concerns about the trustworthiness of model outputs.}
LLMs enable the acquisition of vast and extensive knowledge and have enormous potential to be applied to various tasks. LLMs, such as ChatGPT 1, GPT-4, Claude, and Llama-2 have achieved widespread popularity and adoption across diverse industries and domains. Despite their powerful capabilities, the issue of “hallucination” still poses a concern that LLMs tend to generate inaccurate/fabricated information in generation tasks \cite{cao2023autohall}. Although LLMs can proficiently generate coherent and context-relevant text, they often exhibit a hallucination known as factual hallucination, which seriously weakens the reliability of LLMs in practical applications \cite{zhang2025siren,huang2025survey,li2023trustworthy}. Factual hallucination is one of the least noticeable types of erroneous outputs, because models often express fictional content in a confident tone \cite{jiang2024large}. To explore the differences in the dynamic changes of hidden states in residual flows between successful knowledge recall and failed knowledge stream in the inference process under the hallucination of known facts, Jiang et al. \cite{jiang2024large} collected knowledge query data specifically for this scenario and tested them on a widely used Llama model. Assume the input of T tokens $t_1$, ..., $t_T$, where each token passes through an embedding matrix $E \in \mathbb{R}^{V \times d}$, transforming from vocabulary space to model space. Subsequently, the tokens traverse through L transformer blocks, continuously evolving within the model space, generating a residual stream of shape $T \times L \times d$. Between layer $l - 1$ and $l$, the hidden state $x_i^{l-1}$ of the $i$-th token is updated as follows:
\begin{equation}\label{eq1}
x_i^l = x_i^{l-1} + a_i^l + m_i^l,
\end{equation}
where $a_i^l$ and $m_i^l$ are the outputs from the $l$-th attention and MLP modules.

\textcolor{black}{Hallucinations in LLMs stem from multiple interconnected factors, including imperfections in training data quality such as incomplete coverage, outdated information, or conflicting sources in pre-training corpora, which have been identified as a primary driver of hallucinated outputs in large-scale language models \cite{kalai2025language}. In addition, fundamental aspects of the model’s transformer architecture and autoregressive decoding process further exacerbate this issue. Specifically, attention mechanisms can amplify spurious patterns learned during training, while the next-token prediction objective prioritizes probabilistic fluency and coherence over verifiable factual accuracy, leading to confident generation of ungrounded content even in advanced models \cite{xu2024hallucination}. Empirical analyses further demonstrate that these architectural and probabilistic elements render hallucinations persistent, as models tend to produce plausible guesses when uncertain rather than explicitly expressing knowledge gaps in high-stakes domains such as medicine \cite{alber2025medical}. This tendency is further reinforced by exposure bias in training and evaluation procedures, which implicitly reward complete responses over abstentions, making hallucinations difficult to detect and mitigate in practice \cite{farquhar2024detecting}.}

Because they primarily generate text based on probability, LLMs may create content that does not conform to facts, especially when faced with unknown or ambiguous inputs. This phenomenon may lead users to believe mistakenly in false information, affecting decision-making and behavior. \textcolor{black}{In terms of data security, hallucinations introduce exploitable vulnerabilities: adversaries can leverage carefully crafted adversarial prompts or indirect injections to intentionally trigger or amplify fabricated outputs, facilitating misinformation dissemination, sensitive data inference, or downstream exploitation in integrated systems \cite{omar2025multi}. For example, embedding subtle misleading details in clinical vignettes can cause models to elaborate on false medical facts with high confidence \cite{greshake2023not}, revealing a broader susceptibility of LLMs to adversarial hallucination that persists across different models and settings \cite{zou2023universal}.} This threat is typically the result of inputting misleading information or disruptive data into the model. A conventional classification of hallucination is the intrinsic-extrinsic dichotomy. Intrinsic hallucination occurs when LLM outputs contradict the provided input, such as prompts. On the other hand, extrinsic hallucination occurs when LLM outputs cannot be verified by the information in the input \cite{xu2024hallucination}. According to the study \cite{xu2024hallucination}, hallucination is an inconsistency between commutable LLMs and a \textcolor{black}{computable} ground truth function.
\textcolor{black}{Hallucinations are difficult to eliminate entirely under current training paradigms \cite{xu2024hallucination}, which motivates continued attention to data quality and data-centric evaluation.}

\subsection*{Prompt Leakage}
In the application of LLMs, prompt leakage poses a noteworthy security threat. The leakage of system prompt information may endanger intellectual property rights and serve as adversarial reconnaissance for adversaries \cite{agarwal2024investigating}. Prompt, which can be a question, request, or contextual information, is a text input by a user when interacting with a language model. The model generates corresponding text output based on these prompts. The quality and content of a prompt directly affect the relevance and accuracy of the generated results. Perez \& Ribeiro \cite{perez2022ignore} defined prompt leakage as the behavior of not matching the original target of the prompt with the new target of the printed part or the entire original prompt. Malicious users can attempt prompt \textcolor{black}{leakage} to copy or steal prompts from specific applications, which may be the most crucial part of GPT-3-based applications. Agarwal et al. \cite{agarwal2024investigating} designed a unique threat model and found that LLMs can leak prompt content word for word or explain them based on the threat model. They applied multiple rounds in the threat model and found that it could increase the average Attack Success Rate (ASR) from 17.7 \% to 86.2 \%, causing 99.9 \% leakage to GPT-4 and claude-1.3. LLM sycophancy behavior makes closed and open-source models more susceptible to prompt leakage. Because of the limited effectiveness of existing prompt leaks that mainly rely on manual queries, Hui et al. \cite{hui2024pleak} designed a novel closed box prompt leakage framework (PLeak) to optimize adversarial queries so that when adversaries send them to the target LLM application, the response displays their system prompts. To reconstruct the target system prompt \( p_t \), \( n \) adversarial queries \( q_{\text{adv}}^1, \ldots, q_{\text{adv}}^n \) and a post-processing function \( P \) are crafted. The responses produced by the target LLM application \( f \) for these adversarial queries are aggregated by \( P \) to approximate the original prompt \( p_t \). This process is formulated as follows:
\begin{align}
    p_r &= P(f(q_{\text{adv}}^1), \ldots, f(q_{\text{adv}}^n)) \notag \\
        &= P(f_\theta(p_t \oplus q_{\text{adv}}^1), \ldots, f_\theta(p_t \oplus q_{\text{adv}}^n)), \label{eq3}
\end{align}
where $p_r$ denotes the reconstructed prompt; $f_\theta$ represents the model behavior when the target prompt $p_t$ is perturbed with each adversarial query, and $\oplus$ denotes the combination operation.


\subsection*{Bias}

\textcolor{black}{Bias is treated here as a data-as-source risk arising from skewed, stereotyped, or demographically imbalanced training and feedback data. The following discussion examines how such data shape model behavior and compromise the integrity of generated outputs.}

Generally speaking, LLM conducts training based on large-scale uncorrected Internet data, inherited stereotypes, false statements, derogatory and exclusive language, and other defamation behavior, which have a disproportionate impact on vulnerable and marginalized communities
\cite{bender2021dangers,dodge2021documenting,sheng2021societal}. These harms are called 'social bias,' a subjective and normative term widely used to refer to the differential treatment or outcomes resulting from historical and structural power asymmetry between social groups \cite{gallegos2024bias}. Whether intentional or unintentional, social bias can be expressed through language. Large-scale language models rely on a large amount of text training data, which cannot be managed and validated by a large human collective \cite{navigli2023biases}. Meanwhile, the significant increase in pre-trained corpora makes it difficult to evaluate the features of these data and check their reliability. Thus, the acquired representations may inherit biases and stereotypes present in large text corpora of language, thereby inheriting biases and stereotypes from pre-trained corpora of the internet \cite{liang2021towards}. Therefore, harmful biases such as gender, sexuality, racial bias, and biases related to ethnic minorities and disadvantaged groups may arise \cite{navigli2023biases}. LLMs often use human feedback to fine-tune artificial intelligence assistants. However, human feedback may also encourage models to generate responses based on users' expectations rather than reality. This behavior is called flattery. Artificial intelligence assistants often mistakenly admit their mistakes, provide biased feedback, and imitate user mistakes when questioned. This suggests that flattery is a characteristic of these model training methods \cite{sharma2024towards}.
\textcolor{black}{In other words, biased or preference-skewed feedback data can directly distort the outputs of fine-tuned LLMs.}

Data selection bias is the systematic error resulting from the given selection of text used to train a language model. This bias may occur during the sampling phase when text is recognized or when data is filtered and cleaned \cite{navigli2023biases}. This may lead to or amplify varying degrees of negative social bias. Regarding training data, important context may be overlooked during data collection, and agents used as labels (such as emotions) may incorrectly measure actual outcomes of interest (such as representative harm). Data aggregation may also mask different social groups that should be treated differently, leading to too general models or only representing the majority group \cite{gallegos2024bias}. However, missing contextual data can lead to bias. Even data collected through proper procedures reflects historical and structural biases worldwide.

Notably, with enhanced capabilities, LLMs demonstrate the ability to autonomously infer a wide range of personal author attributes from large volumes of unstructured text provided during inference \cite{staab2024beyond}.
\textcolor{black}{Plant et al. \cite{plant2025you} further showed that pre-trained language models can encode demographic information such as age, gender, and location in text representations.
Such encoded correlations indicate that biased demographic patterns in training data are retained in model behavior and can distort downstream predictions as model scale grows.
From a data-security perspective, this underscores the need to audit and curate training corpora so that spurious demographic signals do not undermine output integrity.}
\textcolor{black}{Related work likewise shows that fine-tuned models may reproduce sensitive attribute correlations learned from training data, which can further amplify biased or unreliable generations if the underlying corpora are not properly controlled.}

\subsection*{\textcolor{black}{Adaptive and Multi-stage Attacks}}

\textcolor{black}{As outlined in the attacker model above, these attacks combine capabilities across training, retrieval, and inference rather than exploiting a single pipeline stage. The following discussion focuses on coordinated poisoning and adaptive evasion strategies.}

\textcolor{black}{Beyond the threats discussed above, recent research has identified a new generation of advanced adversarial threats that exploit the architectural complexity and multi-stage inference pipelines of modern LLMs \cite{jiang2025safety}.}
\textcolor{black}{Beyond single-attacker scenarios, knowledge-base poisoning attacks can inject multiple adversarial texts into retrieval corpora to manipulate generation while evading detection systems that rely on per-document screening \cite{zou2025poisonedrag}.} 
\textcolor{black}{ Attackers can further employ stealthy and adaptive strategies, where they actively monitor defense mechanisms and adjust their attack patterns in response; such adversaries have been shown to maintain high success rates even against adversarially trained models by iteratively refining their attacks based on model feedback \cite{tramer2020adaptive}.}
\textcolor{black}{ In such multi-stage settings, poisoned data may first enter a training corpus or retrieval index and later be activated through prompts, retrieved context, or tool use, making these threats difficult to isolate with defenses deployed at a single stage.}

\subsection*{\textcolor{black}{RAG and Agent Security Risks}}

\textcolor{black}{As outlined in the attacker model above, these risks arise once external corpora, vector indexes, tools, memory, and inter-agent messages become part of the model context. The following discussion covers retrieval-stage and agent-specific attack surfaces beyond standalone chat models.}

\textcolor{black}{In RAG systems, such backdoors can be planted directly into retriever embeddings or external knowledge bases, making triggered documents preferentially retrieved and subsequently influencing generation in ways that evade standard detection \cite{zeng2024good}.}
\textcolor{black}{ This pattern also covers vector and database poisoning, in which adversaries manipulate embedding stores or knowledge bases so that malicious documents are preferentially retrieved at inference time.}
\textcolor{black}{ Indirect prompt injection is a closely related risk: attacker-controlled text in webpages, emails, or retrieved documents may be interpreted as instructions rather than inert content once incorporated into the model context \cite{greshake2023not}.}
\textcolor{black}{Retrieval pipelines further create leakage risks when sensitive information can be extracted from RAG datastores via prompt-injected queries \cite{qi2025follow}.}

\textcolor{black}{The proliferation of LLM-based agents and tool-augmented systems has introduced entirely new attack surfaces, as malicious instructions embedded in untrusted external content can manipulate tool-integrated agents into executing harmful or unauthorized actions \cite{zhan2024injecagent}.}
\textcolor{black}{ Agent deployments extend this attack surface to tool-output poisoning, memory poisoning, unauthorized tool use, and cross-session leakage through persistent agent state \cite{zhan2024injecagent,chen2024agentpoison}.}
\textcolor{black}{ Inter-agent attacks allow malicious instructions to propagate across multi-step workflows, while compromised tool outputs can redirect agents toward unintended actions or data exfiltration. Tool-augmented attacks represent an especially severe threat, enabling adversaries to trick LLMs into invoking external tools, such as search engines, code executors, or database queries in unintended ways that facilitate data exfiltration or system compromise \cite{greshake2023not}.}
\textcolor{black}{ These advanced threats collectively reveal that static, isolated defenses are insufficient, necessitating mitigation strategies that account for distributed, adaptive, and multi-agent attack paradigms across the entire LLM lifecycle \cite{zhang2020adversarial}.}

\textcolor{black}{\subsection*{Real-World Deployment Considerations}}

\textcolor{black}{Understanding data security risks in isolation is insufficient; their practical impact depends heavily on the deployment context of LLMs. For training-time risks such as data poisoning and bias, the primary concern is the supply chain of training data. In industrial settings where models are continuously fine-tuned on user feedback or external data sources, the attack surface for poisoning expands significantly, requiring rigorous data provenance and sanitization pipelines before training. \textcolor{black}{A representative case is the web-scale poisoning study by Carlini et al. \cite{carlini2024poisoning}, which showed that popular training datasets can be poisoned through split-view poisoning and frontrunning attacks, exposing the fragility of large-scale data supply chains used for modern model training.} For inference-time risks such as prompt injection, jailbreak, and prompt leakage, deployment architecture determines vulnerability exposure. Cloud-based LLM APIs with shared system prompts face higher prompt leakage risks, while locally deployed models may be more susceptible to jailbreak attempts due to limited monitoring infrastructure \cite{greshake2023not}. \textcolor{black}{Another practical case is the indirect prompt injection study by Greshake et al. \cite{greshake2023not}, which demonstrated attacks against real-world LLM-integrated applications, including Bing Chat and code-completion systems, where attacker-controlled webpages or code comments could be processed as instructions and used for behavior manipulation or data exfiltration.} RAG-integrated systems introduce indirect prompt injection vectors through retrieved documents, necessitating additional filtering at the retrieval stage \cite{liu2023prompt}. For hallucination, acceptable tolerance varies by application domain. In high-stakes domains such as healthcare or finance, even low-rate hallucinations can lead to severe consequences, requiring domain-specific mitigation strategies \cite{alber2025medical}. In contrast, creative applications may tolerate higher hallucination rates. Deployment decisions must therefore consider risk-utility trade-offs specific to each use case.}


\begin{table}[!htbp]
\centering
\color{black}
\footnotesize
\setlength{\tabcolsep}{3.5pt}
\caption{\textcolor{black}{data-centric taxonomy of security risks in LLMs.}}
\label{tab:data_taxonomy}
\begin{tabularx}{\linewidth}{@{}%
>{\RaggedRight\arraybackslash}p{0.22\linewidth}%
>{\RaggedRight\arraybackslash}p{0.26\linewidth}%
>{\RaggedRight\arraybackslash}p{0.20\linewidth}%
>{\RaggedRight\arraybackslash}X%
@{}}
\toprule
\textbf{Threat} & \textbf{Attack Lifecycle} & \textbf{Data Role} & \textcolor{black}{\textbf{Dimension}} \\
\midrule
Data Poisoning & Training-time / Fine-tuning & Data as target & \textcolor{black}{Threat mechanism} \\
Prompt Injection & Inference-time / Retrieval-stage & Data as vector & \textcolor{black}{Threat mechanism} \\
Jailbreak & Inference-time & Data as vector & \textcolor{black}{Threat mechanism} \\
Hallucination & Training-time / Inference-time & Data as source & \textcolor{black}{Threat mechanism} \\
Prompt Leakage & Inference-time / Agent-based & Data as target & \textcolor{black}{Threat mechanism} \\
Bias & Training-time / Fine-tuning & Data as source & \textcolor{black}{Threat mechanism} \\
Adaptive and Multi-stage Attacks & Retrieval-stage / Multi-stage & Data as target / vector & \textcolor{black}{Attacker strategy} \\
RAG and Agent Security Risks & Retrieval-stage / Agent-based & Data as target / vector & \textcolor{black}{Deployment surface} \\
\bottomrule
\end{tabularx}
\end{table}

\textcolor{black}{Table~\ref{tab:data_taxonomy} summarizes the risks discussed above along attack lifecycle and data role, and introduces a dimension column to clarify the perspective represented by each category.
Poisoning, prompt injection, jailbreak, hallucination, prompt leakage, and bias are listed as threat mechanisms; among them, hallucination and bias are included specifically as data-as-source risks.
Adaptive and multi-stage attacks are better viewed as an attacker strategy, while RAG and agent security risks are better viewed as a deployment surface.
Lifecycle and data-role labels remain complementary and are not mutually exclusive, since a single incident may span multiple stages or combine more than one data role.
This presentation keeps the taxonomy data-centric while making overlaps among mechanism, strategy, and deployment surface easier to interpret.}

\begin{table*}[tp]
\renewcommand{\arraystretch}{1.4} 
\setlength{\tabcolsep}{7pt}     
\Huge
\centering
\caption{\textcolor{black}{Strategies for protecting data security. This table categorizes representative defense methods for LLM security, including training data cleaning, adversarial training, output guardrails, RLHF, data augmentation, and RAG/Agent Defenses (corresponding to the subsection Deployment-specific defenses for RAG and agents). For each approach, it lists the techniques used, tested models, benchmark datasets, and evaluation metrics from relevant studies.}}
\resizebox{1.05\linewidth}{!}{
\begin{tabular}{lllllll}
\hline
\multicolumn{1}{l}{Category} & \multicolumn{1}{l}{Work} & \multicolumn{1}{l}{Method} & \multicolumn{1}{l}{Evaluated Model} & \multicolumn{1}{l}{Dataset} & \multicolumn{1}{l}{Evaluation Metric} & \multicolumn{1}{l}{\textcolor{black}{Defended Against}} \\ \hline

& \textcolor{black}{\cite{li2026backdoorllm}} & \textcolor{black}{Backdoor attacks} & \begin{tabular}[c]{@{}l@{}}\textcolor{black}{LLaMA-7B, LLaMA-13B, }\\ \textcolor{black}{ LLaMA-70B} \end{tabular} & \begin{tabular}[c]{@{}l@{}}\textcolor{black}{Stanford Alpaca} \\ \textcolor{black}{AdvBench} \end{tabular} & \textcolor{black}{ASR} & \\ 

\begin{tabular}[c]{@{}l@{}}\textcolor{black}{Training Data} \\ \textcolor{black}{Cleaning} \end{tabular} & \textcolor{black}{\cite{souly2025poisoning}} & \textcolor{black}{Empirical poisoning} & \textcolor{black}{Pythia 6.9B} & \begin{tabular}[c]{@{}l@{}}\textcolor{black}{Chinchilla-optimal} \\ \textcolor{black}{pretraining corpora} \end{tabular} & \begin{tabular}[c]{@{}l@{}}\textcolor{black}{ASR, CA}\\ \textcolor{black}{NTA} \end{tabular} &  \begin{tabular}[c]{@{}l@{}}\textcolor{black}{Data Poisoning,}\\ \textcolor{black}{Adaptive and}\\ \textcolor{black}{Multi-stage Attacks}\end{tabular} \\ 

& \textcolor{black}{\cite{fu2024poisonbench}} & \begin{tabular}[c]{@{}l@{}}\textcolor{black}{Content injection} \\ \textcolor{black}{Alignment deterioration} \end{tabular} & \textcolor{black}{GPT-3.5, GPT-4 / GPT-4o, etc } & \begin{tabular}[c]{@{}l@{}}\textcolor{black}{Anthropic HH-RLHF, }\\ \textcolor{black}{Ultrafeedback} \end{tabular} & \textcolor{black}{AS, SS} & \\ \hline

& \cite{han2024evaluation} & Automated Injection & ChatGPT & / & \begin{tabular}[c]{@{}l@{}}Offensive Language Detection, etc\end{tabular} & \\ 

& \cite{paudice2018detection} & \begin{tabular}[c]{@{}l@{}}Adversarial machine \\ learning \end{tabular} & linear models & Spambase, MNIST & Classification Error & \\  

Adversarial Training & \cite{geiping2021doesn} & Deep neural networks & GPT-3.5 / GPT-4 & GTSRB, CIFAR-10 & Acc & \textcolor{black}{Prompt Injection} \\

& \cite{wen2023adversarial} & Adversarial training & / & \begin{tabular}[c]{@{}l@{}}CIFAR-10, CIFAR-100,\\TINYIMAGENET\end{tabular} & Acc & \\

& \cite{tramer2020adaptive} & AutoAttack & \begin{tabular}[c]{@{}l@{}}Madry’s PGD-trained ResNe, etc\end{tabular} & CIFAR-10, ImageNet, MNIST & Robust accuracy & \\ \hline

& \textcolor{black}{\cite{han2024wildguard}} & \textcolor{black}{WildGuardMix} & \textcolor{black}{GPT-4} & \begin{tabular}[c]{@{}l@{}}\textcolor{black}{WildGuardTrain,}\\ \textcolor{black}{WildGuardTest}\end{tabular} & \begin{tabular}[c]{@{}l@{}}\textcolor{black}{F1 score,}\\ \textcolor{black}{Jailbreak success reduction}\end{tabular} & \\   

\textcolor{black}{Output \textcolor{black}{Guardrails}} & \textcolor{black}{\cite{zeng2024autodefense}} & \begin{tabular}[c]{@{}l@{}}\textcolor{black}{Multi-agent }\\ \textcolor{black}{response-filtering}\end{tabular} & \textcolor{black}{GPT-3.5} & \begin{tabular}[c]{@{}l@{}}\textcolor{black}{Harmful and }\\ \textcolor{black}{safe prompts}\end{tabular} & \begin{tabular}[c]{@{}l@{}}\textcolor{black}{ASR, FPR,}\\ \textcolor{black}{Normal Request Accuracy}\end{tabular} &  \begin{tabular}[c]{@{}l@{}}
   \textcolor{black}{Jailbreak,}\\
   \textcolor{black}{RAG and Agent}\\
   \textcolor{black}{Security Risks}
   \end{tabular}\\   

& \textcolor{black}{\cite{ghosh2025aegis2}} & \textcolor{black}{Hybrid data curation} & \textcolor{black}{Llama-3.1-8B-Instruct} & \textcolor{black}{Aegis 2.0 Dataset} & \begin{tabular}[c]{@{}l@{}}\textcolor{black}{Classification Accuracy,}\\ \textcolor{black}{Hazard Category F1, etc}\end{tabular} & \\ \hline

& \cite{ouyang2022training} & \begin{tabular}[c]{@{}l@{}}Supervised learning,\\ RL \end{tabular} & GPT-3 & SFT, RM, PPO & human preference ratings & \\ 

\begin{tabular}[c]{@{}l@{}}RLHF\end{tabular} & \cite{yu2024rlhf} & \begin{tabular}[c]{@{}l@{}}Dense Direct Preference \\ Optimization\end{tabular} & \begin{tabular}[c]{@{}l@{}}Muffin, LRV, etc\end{tabular} & / & \begin{tabular}[c]{@{}l@{}}Object HalBench, \\ MMHal-Bench, etc\end{tabular} & \textcolor{black}{Hallucination} \\      

& \cite{christiano2017deep} & Deep neural networks & reward model & Atari, MuJoCo & reward & \\      

& \cite{burns2024discoveringlatentknowledgelanguage} & Linear probe & GPT-2, GPT-J & \begin{tabular}[c]{@{}l@{}}Wikidata-derived \\ factual triples\end{tabular} & \begin{tabular}[c]{@{}l@{}}Probe Accuracy, \\ Precision, KL divergence\end{tabular} & \\ \hline

& \cite{tokpo2024fairflow} & \begin{tabular}[c]{@{}l@{}}CDA, Disentangling invertible, \\ Interpretation network\end{tabular} & \begin{tabular}[c]{@{}l@{}}BART, ChatGPT, \\ FairFlowV2, etc\end{tabular} & Bias-in-bios, ECHR, Jigsaw & Acc, PPL, F1, FPRD, TPRD & \\ 

Data Augmentation & \cite{maudslay2019s} & \begin{tabular}[c]{@{}l@{}}CDA, Names Intervention\end{tabular} & / & \begin{tabular}[c]{@{}l@{}}SSA, SimLex-999,\\ Doc2Vec\end{tabular} & Error rate & \begin{tabular}[c]{@{}l@{}}\textcolor{black}{Prompt Leakage,}\\ \textcolor{black}{Bias}\end{tabular} \\                                                                                      

& \cite{yu2023mixup} & \begin{tabular}[c]{@{}l@{}}Natural language \\ processing\end{tabular} & SOTA & TREC, AG’s News & SEAT, Acc & \\ \hline

& \textcolor{black}{\cite{zou2025poisonedrag}} &
\textcolor{black}{\begin{tabular}[c]{@{}l@{}}Corpus / index\\ integrity controls\end{tabular}} &
\textcolor{black}{\begin{tabular}[c]{@{}l@{}}GPT-4, Llama2-7B,\\ Vicuna-7B, etc\end{tabular}} &
\textcolor{black}{\begin{tabular}[c]{@{}l@{}}NQ, HotpotQA,\\ MS-MARCO\end{tabular}} &
\textcolor{black}{ASR} & \\ 

\textcolor{black}{\begin{tabular}[c]{@{}l@{}}RAG/Agent \\ Defenses\end{tabular}} &
\textcolor{black}{\cite{qi2025follow}} &
\textcolor{black}{\begin{tabular}[c]{@{}l@{}}Authorization-aware\\ retrieval controls\end{tabular}} &
\textcolor{black}{\begin{tabular}[c]{@{}l@{}}Llama2, Mistral,\\ Vicuna, GPTs, etc\end{tabular}} &
\textcolor{black}{\begin{tabular}[c]{@{}l@{}}Customized GPTs,\\ Enron email, book corpus\end{tabular}} &
\textcolor{black}{\begin{tabular}[c]{@{}l@{}}Verbatim\\ extraction rate\end{tabular}} &
\textcolor{black}{\begin{tabular}[c]{@{}l@{}}RAG and Agent\\ Security Risks\end{tabular}} \\ 

& \textcolor{black}{\cite{zhan2024injecagent}} &
\textcolor{black}{\begin{tabular}[c]{@{}l@{}}Tool / memory runtime\\ constraints\end{tabular}} &
\textcolor{black}{\begin{tabular}[c]{@{}l@{}}GPT-4, GPT-3.5,\\ Llama2-70B, etc\end{tabular}} &
\textcolor{black}{\begin{tabular}[c]{@{}l@{}}InjecAgent benchmark\\ (1,054 test cases)\end{tabular}} &
\textcolor{black}{ASR-valid} & \\ \hline

\end{tabular}}
\label{tab:example4}
\end{table*}

\section*{Defense Strategies}


To combat the various threats to data security, a range of defense strategies has been proposed (as shown in Table~\ref{tab:example4}). \textcolor{black}{Rather than aiming to cover every existing method, this section organizes representative and widely evaluated defense families that recur across the LLM data-security literature, including training data cleaning, adversarial training, output moderation and guardrails, RLHF, data augmentation, and RAG/agent defenses.} We classify these strategies and review their main mechanisms, strengths, and limitations in the following subsections.

\subsection*{Training Data Cleaning}

\textcolor{black}{Training data cleaning techniques focus on preemptively identifying and removing suspicious or poisoned samples from datasets to mitigate training-time vulnerabilities, such as data poisoning and backdoor attacks. These attacks can be highly effective even with minimal contamination. For example, poisoning just a small fraction of the training data can implant persistent backdoors that trigger malicious behaviors while preserving performance on benign evaluations. Recent studies demonstrate that as few as 250 poisoned documents can successfully embed backdoors in models ranging from 600M to 13B parameters during pretraining, with the required number of poisons remaining nearly constant regardless of dataset scale \cite{souly2025poisoning, li2026backdoorllm}. In sensitive domains like medicine, injecting misinformation into merely 0.001\% of tokens has been shown to significantly increase the likelihood of harmful outputs (e.g., by 7 to 11\%) without degrading clean benchmark performance \cite{alber2025medical}.}

\textcolor{black}{Defensive approaches include anomaly detection to flag outliers, provenance tracking to verify data origins, and knowledge graph validation to ensure factual consistency, particularly effective in biomedical contexts \cite{zhou2025survey}. \textcolor{black}{For knowledge-graph-based sanitization, reproducible implementation should specify the knowledge graph source, entity-linking rules, relation-matching criteria, and the rule used to reject unsupported or contradictory claims. In biomedical settings, such validation is particularly relevant because poisoned medical misinformation can remain hidden under standard benchmark evaluation while still increasing harmful outputs \cite{alber2025medical}.} Specialized benchmarks such as BackdoorLLM \cite{li2026backdoorllm} and PoisonBench \cite{fu2024poisonbench} provide standardized evaluations. These benchmarks reveal that methods like noise-augmented contrastive learning, frequency-space downscaling, and contrastive purification can substantially reduce backdoor persistence while maintaining model utility. These cleaning strategies are especially critical for large-scale pretraining pipelines, where exhaustive manual inspection is infeasible. They form a foundational layer in multi-stage defenses against supply-chain vulnerabilities.}

\textcolor{black}{Beyond these passive filtering approaches, recent advances in dataset purification have introduced proactive methods that leverage generative model dynamics to actively cleanse poisoned samples. For instance, energy-based models (EBMs) and denoising diffusion probabilistic models (DDPMs) can apply stochastic transformations to purify contaminated data without requiring attack-specific prior knowledge, effectively sanitizing inputs before they reach the training pipeline \cite{bhat2024puregen}. \textcolor{black}{For purification methods such as PureGen, reproducibility depends on reporting the purifier type, the corruption or diffusion noise schedule, the number of purification steps, and the criterion used to decide whether a sample is sanitized or discarded \cite{bhat2024puregen}. These settings determine both sanitization strength and computational cost.} \textcolor{black}{This type of purification is most suitable under training-time or fine-tuning threat models where defenders can process data before model updating, because PureGen is designed as a pre-training data purification step rather than an inference-time guardrail \cite{bhat2024puregen}. Its main advantage is that it does not rely on knowing the exact trigger or poisoning pattern, but the additional generative preprocessing required for each suspicious sample limits its scalability for web-scale pre-training corpora. Therefore, it is more practical for curated fine-tuning datasets, domain-specific corpora, or high-risk data subsets than for exhaustive cleaning of all pre-training data.}}

\subsection*{Adversarial Training}
Adversarial training desensitizes neural networks to adversarial perturbations in testing time by adding temporary adversarial examples to the training data \cite{mkadry2017towards}. The purpose of adversarial training is to improve the security and robustness of LLMs through the use and training of adversarial samples, enabling the models to better cope with various challenges that may be encountered in reality. 

The study found valuable insights into the vulnerability of LLMs such as ChatGPT when subjected to malicious prompt injection. The identification of significant rates of harmful reactions in various situations highlights the need for continuous research and development to improve safety and reliability; whereas, advanced adversarial training techniques expose models to a wide range of adversarial inputs and enhance their resilience \cite{han2024evaluation}. Coincidentally, data poisoning refers to adversaries disrupting the learning process by injecting malicious samples into the training data \cite{paudice2018detection}. At present, various defense measures have been proposed for the threat model of data poisoning; however, each defense measure has different shortcomings, such as being easily overcome by adaptive attacks, seriously reducing testing performance, or being unable to be generalized to various data poisoning threat models. Adversarial training and its variants are currently judged to be the only empirically strong defense against (inference-time) adversarial attacks \cite{geiping2021doesn}. Even so, throughout the training process of Wen et al. \cite{wen2023adversarial}, the adversarial risks of clean data and toxic data confirmed their claim that adversarial training faces difficulties in optimizing toxic data because the speed of risk reduction is slower than in clean situations. Adversarial training also solves the following saddle-point problem:
\begin{equation}\label{eq4}
\mathop{\min}_{\theta} \mathbb{E}_{(x,y)\sim\mathbb{D}} \big[\max_{\Delta\in S} \mathcal{L}_{\theta} (\mathnormal{x} + \Delta, \mathnormal{y})\big],
\end{equation}
where $\mathcal{L}_{\theta}$ denotes the loss function of a model with parameters $\theta$, and the adversary perturbs inputs
$x$ from a data distribution $\mathbb{D}$, subject to the constraint that perturbation $\Delta$ is in $S$ \cite{geiping2021doesn}. Geiping et al. \cite{geiping2021doesn} proposed a variant of adversarial training that uses adversarial poisoning data instead of adversarial examples during testing, thereby modifying the training data to desensitize the neural network to the types of perturbations caused by data poisoning. 
\textcolor{black}{For reproducibility, adversarial training should report the perturbation constraint set $S$, the poisoning budget or poison ratio, the number of inner maximization steps, the step size, and the projection or clipping rule used to keep adversarial samples within the allowed perturbation bound \cite{geiping2021doesn}. These parameters directly determine the strength of the generated adversarial samples, the clean-accuracy trade-off, and the computational cost of training.} 
\textcolor{black}{This defense is most suitable when defenders can anticipate the poisoning mechanism and generate representative adversarial poisoning samples before training. It directly improves robustness against the perturbations used during training, but generalizes poorly to unseen or adaptive poisoning strategies. Wen et al. \cite{wen2023adversarial} show that adversarial training may optimize toxic data more slowly than clean data, so robustness gains can be uneven when poisoned and benign samples coexist. Adaptive attack evaluations further suggest that robustness may not transfer reliably across threat models, and adversarial training should therefore not be treated as a universal defense \cite{tramer2020adaptive}. Because generating adversarial samples and repeatedly updating model parameters is computationally expensive, this approach is more practical for fine-tuning, smaller safety models, or controlled experiments than for repeated training of frontier-scale LLMs.}

\subsection*{Output \textcolor{black}{Guardrails}}

\textcolor{black}{Output \textcolor{black}{guardrails} and moderation involve post-generation filtering to inspect and block harmful, biased, or leaked content in model responses. This approach addresses inference-time risks including jailbreaks and policy violations. \textcolor{black}{This mechanism is particularly critical against \textcolor{black}{RAG and agents security risks} such as indirect prompt injection attacks that embed malicious instructions within retrieved documents to manipulate model behavior \cite{liu2023prompt}.} Advances in 2024 to 2025 emphasize hybrid guardrails that integrate rule-based filters, toxicity classifiers, and dedicated moderation models for real-time detection. \textcolor{black}{For reproducible moderation, deployments should specify whether filtering is applied to prompts, responses, or both, which risk categories are enabled, the blocking threshold for unsafe content, and the refusal template or fallback action used when a sample is rejected. These sanitization rules are essential because different thresholds can trade off jailbreak resistance against false positives and excessive refusals.} For instance, WildGuard, an open-source multi-task classifier, simultaneously evaluates prompt maliciousness, response harmfulness, and refusal appropriateness across 13 risk categories. It outperforms prior open models. It approaches proprietary systems like GPT-4 in accuracy while reducing jailbreak success rates from ~80\% to under 3\% when deployed as an interface moderator \cite{han2024wildguard}. \textcolor{black}{WildGuard is therefore most suitable for black-box inference-time threat models, where the protected LLM cannot be retrained or inspected. Its main advantage is deployment scalability, since a separate moderation model can be attached to different LLM applications; however, classifier-based guardrails may still produce false positives and can be bypassed by adaptive prompts that are optimized against the deployed detector.}}

\textcolor{black}{Multi-agent frameworks, such as AutoDefense, distribute defense roles across LLM agents for collaborative filtering, enhancing robustness against diverse jailbreaks \cite{zeng2024autodefense}. \textcolor{black}{For reproducibility, AutoDefense should report the number of agents, the base model used by each agent, the role-specific safety prompts, and the aggregation rule used to decide whether a response should be blocked or replaced with an explicit refusal.} Constitutional classifiers, which enforce alignment via principle-based decision rules, have proven effective against universal adversarial prompts, maintaining low false positives \cite{sharma2025constitutional}. Evaluations on comprehensive safety benchmarks and diverse risk taxonomies highlight the advantages of hybrid approaches, combining specialized classifiers with LLM-as-judge mechanisms, to address emerging threats including indirect prompt injections and subtle harms, albeit with persistent challenges in managing latency, excessive refusals, and generalization to novel risks \cite{ghosh2025aegis2}. \textcolor{black}{AutoDefense is particularly effective when the threat model involves diverse jailbreak prompts and the defender can afford multiple safety-checking steps before returning a response. Under the DAN jailbreak setting with GPT-3.5 Turbo as the victim model, AutoDefense reduces ASR from 55.74\% without defense to 7.95\% when using LLaMA-2-13B in a three-agent defense system, illustrating the benefit of decomposing moderation into specialized agent roles \cite{zeng2024autodefense}. However, this design increases inference-time cost because each user request requires several additional model calls, making latency and serving cost important limitations for real-time or high-throughput deployments.}}

\subsection*{Reinforcement Learning From Human Feedback}

When LLMs become larger and more complex, they may output incorrect and useless content to users, leading to hallucinations. Nonetheless, reinforcement learning or fine-tuning of the model through human feedback can solve or weaken such phenomena \cite{ouyang2022training}. RLHF optimizes the model by combining human feedback to make its output more in line with human expectations. With the intention of aligning with human preferences, RLHF typically employs reinforcement learning algorithms to optimize LLMs, generating outputs that maximize the rewards provided by training preference models. Besides, integrating human feedback into the training cycle of LLMs can enhance their consistency and guide them to produce high-quality and harmless responses \cite{huang2025survey}.

Based on the fact that existing Multimodal LLMs commonly suffer from severe hallucinations and generate text that is not based on relevant content, Yu et al. \cite{yu2024rlhf} proposed RLHF-V to address this issue. In particular, RLHF-V collects human preferences in the form of fragment-level hallucination correction and performs intensive direct preference optimization on human feedback. The comprehensive experiments have shown that RLHF-V can greatly improve the credibility of LLMs in generating good data and computational efficiency. \textcolor{black}{RLHF-V is most suitable for multimodal hallucination threat models where erroneous textual descriptions can be localized and corrected through human feedback. Its fragment-level correction provides more fine-grained supervision than response-level preference labels, but this also increases annotation cost and limits scalability when applied to large and diverse multimodal datasets \cite{yu2024rlhf}. Moreover, because RLHF-V focuses on hallucination correction rather than adversarial manipulation, it should be viewed as a reliability-oriented alignment method rather than a standalone defense against jailbreaks, prompt injection, or data poisoning.} Over the long term, learning tasks from human preferences is no more difficult than learning tasks from programmatic reward signals, ensuring that powerful reinforcement learning systems can be applied to complex human values rather than low complexity goals \cite{christiano2017deep}.

\subsection*{Data Augmentation}

Data augmentation techniques mitigate or eliminate bias by adding new examples to the training data. These examples increase the diversity and quantity of the training dataset, thereby expanding the distribution of underrepresented or misrepresented social groups, which can then be used for training \cite{gallegos2024bias}. This exposes the model to a wider and more balanced data distribution during training. 

Counterfactual Data Augmentation (CDA), one of the main techniques in data augmentation technology, aims to balance the demographic attributes in training data and has been adopted widely to mitigate bias in NLP \cite{tokpo2024fairflow}. Conversely, due to the potential quality problems of this technology and the high cost of data collection, Tokpo \& Calders \cite{tokpo2024fairflow} proposed FairFlow, a method for automatically generating parallel data for training counterfactual text generator models that limit the need for human intervention. FairFlow can significantly overcome the limitations of dictionary-based word replacement methods while maintaining good performance. \textcolor{black}{FairFlow is most suitable for bias-related fine-tuning settings where demographic attributes can be explicitly identified and counterfactual pairs can be generated before model updating. Its advantage is that it reduces the need for manual counterfactual annotation, but its effectiveness still depends on the quality of generated parallel data and may be limited when demographic attributes are implicit, intersectional, or context-dependent \cite{tokpo2024fairflow}.} As for the part of model training (fine-tuning) in the entire method, the approach involves fine-tuning a BART model on the parallel data generated from previous steps. The BART generator takes the original source text \( X \) as input and is trained to autoregressively generate the counterfactual text \( Y \), using the corresponding counterfactual references as supervision in a teacher-forcing manner. This objective can be formulated as follows:
\begin{equation}\label{eq5}
\mathcal{L}_{\text{generator}} = - \sum_{t=1}^{k} \log P(y_t \mid Y_{<t}, X),
\end{equation}
where \( X \) and \( Y \) represent the source and target texts, respectively. Here, \( y_t \in Y \) denotes the \( t^{\text{th}} \) token in the target text, and \( Y_{<t} \) refers to all tokens in \(Y\) preceding \(y_t\).
Maudslay et al. \cite{maudslay2019s} made two improvements to CDA: one, Counterfactual Data Substitution (CDS), is a variant of CDA in which potentially biased text is randomly replaced to avoid duplication. The other, name intervention, can deal with the inherent bias of names. Name intervention adopts a novel name-pairing strategy that takes into account both the frequency of the name and the gender specificity.

\begin{figure}[H]
  \centering
  \includegraphics[width=\textwidth]{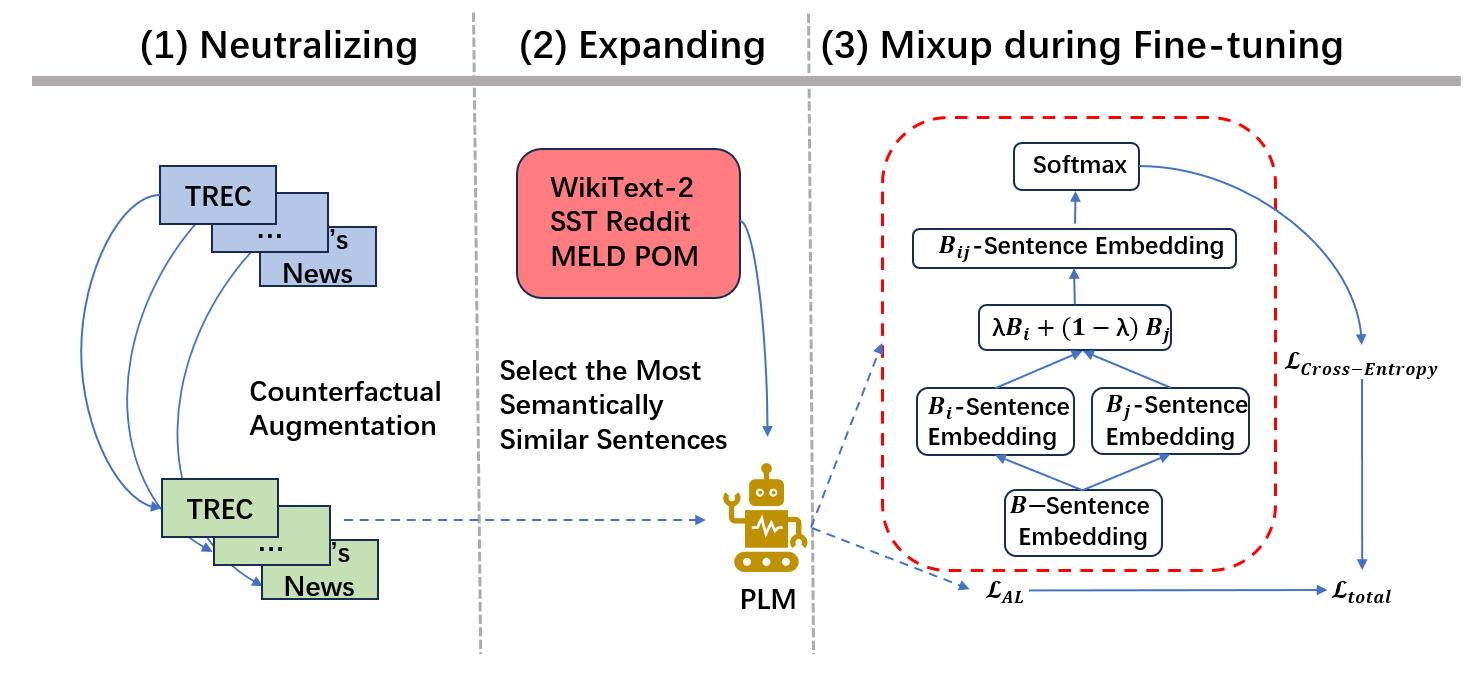}
  \caption{\textcolor{black}{The overall architecture of Mix-Debias, a three-stage, training-time defense against bias in fine-tuning data. \textbf{Stage~1 (Neutralizing):} counterfactual data augmentation replaces stereotype-linked attributes to reduce spurious correlations in task datasets (e.g., TREC, AG's News). \textbf{Stage~2 (Expanding):} a PLM selects semantically similar sentences from external corpora (WikiText-2, SST, Reddit, etc.) to enrich the debiased dataset. \textbf{Stage~3 (Mixup):} $\lambda$-weighted interpolation of sentence embeddings during fine-tuning promotes smoother decision boundaries and reduces memorization of biased patterns. This framework mitigates the influence of biased or contaminated training data before it is encoded into model parameters, but it does not directly address inference-time threats such as prompt injection or jailbreak~\cite{yu2023mixup}.}}
  \label{fig4}
\end{figure}

To remove the undesired stereotyped associations in models during fine-tuning, Yu et al. \cite{yu2023mixup} proposed a mixture-based framework (Mix-Debias) from a new unified perspective, which directly combines the debiased models with fine-tuning applications. Mix-Debias applies CDA to obtain gender-balanced correspondence of downstream task datasets. Then, it further selects the most semantically meaningful sentences from a rich additional corpus to expand the previously neutralized dataset. \textcolor{black}{Compared with simple CDA, Mix-Debias is more scalable for downstream fine-tuning because it augments counterfactual examples with additional semantically relevant data rather than relying solely on attribute substitution. However, it mainly targets bias mitigation rather than adversarial security threats, and its benefits may not extend to prompt injection, jailbreak, or retrieval-stage attacks introduced after training \cite{yu2023mixup}.} The overall architecture of Mix-Debias is illustrated in Fig.~\ref{fig4}.

\subsection*{Deployment-specific defenses for RAG and agents}

\textcolor{black}{The defense families reviewed above remain useful in RAG and agent settings, but they do not by themselves cover retrieval corpora, vector indexes, tools, or persistent memory. The RAG and agent risks surveyed earlier show that external documents and tool outputs can become attack vectors once they enter the model context \cite{greshake2023not}. The controls below are therefore best viewed as deployment-specific complements to training data cleaning, adversarial training, output guardrails, RLHF, and data augmentation, rather than as replacements for them.}

\textcolor{black}{For RAG pipelines, corpus provenance and document sanitization extend the integrity concerns already raised for training-data supply chains to external knowledge stores. Provenance documentation helps establish where retrieved documents come from and how they influence model behavior \cite{kale2023provenance}. Source verification likewise matters because web-scale data pipelines can be poisoned when crawled content is not checked against trusted origins \cite{carlini2024poisoning}. Document sanitization is further motivated by safety analyses showing that corrupted or unsafe retrieved content can degrade RAG behavior \cite{an2025rag}. Authorization-aware retrieval is especially relevant when prompt-injected queries can solicit sensitive datastore content that should not be returned for a given user or task \cite{qi2025follow}. Privacy-oriented RAG studies also show that embeddings and vector indexes themselves are sensitive assets that require access control \cite{zeng2024good}. Instruction--data separation follows directly from indirect prompt injection: retrieved text should be treated as untrusted content rather than executable instructions \cite{greshake2023not}. Retrieved-document prompt-injection detection should therefore be applied at the retrieval or context-construction stage before generation \cite{liu2023prompt}. Vector-store poisoning detection is motivated by knowledge-corruption attacks that manipulate retrieval corpora or indexes so that malicious documents are preferentially retrieved \cite{zou2025poisonedrag}. In practice, these RAG controls can be paired with output guardrails when malicious retrieved content still reaches the generator \cite{han2024wildguard}. Multi-agent response filtering provides an additional layer when a single moderator is insufficient \cite{zeng2024autodefense}.}

\textcolor{black}{For agent deployments, the main goal is to constrain the operational impact of untrusted tool outputs and persistent state. Tool sandboxing and least-privilege access are natural responses to demonstrations that tool-integrated workflows can be coerced into unauthorized tool use \cite{zhan2024injecagent}. Indirect prompt injection in LLM-integrated applications likewise shows why untrusted external content must not be granted unrestricted tool authority \cite{greshake2023not}. Approval of high-risk actions adds a human or policy gate before irreversible operations of the kind examined in agent injection benchmarks \cite{zhan2024injecagent}. Memory quarantine is motivated by memory- and knowledge-base poisoning in agent systems, where compromised state can persist across steps or sessions \cite{chen2024agentpoison}. Audit logs make unauthorized tool use reconstructible after the fact \cite{zhan2024injecagent}. Secret redaction reduces exposure of sensitive datastore content or tool returns in agent traces \cite{qi2025follow}. Rollback mechanisms limit the lasting effect of poisoned memory or corrupted intermediate state \cite{chen2024agentpoison}.}

\textcolor{black}{Overall, RAG and agent defenses should be layered with the generic methods in this section: cleaning and provenance protect corpora and indexes, guardrails and injection detectors inspect retrieved or tool-derived text, and runtime controls restrict what agents may do when those filters fail. No single control covers the full retrieval--generation--action pipeline, which is consistent with the multi-stage nature of the surveyed threats \cite{zou2025poisonedrag}.}

{\textcolor{black}{\subsection*{Real-World Application Considerations}}

\textcolor{black}{While the defense strategies surveyed above achieve promising results in experimental settings, several practical constraints shape their effectiveness in production deployments. \textcolor{black}{Most defenses are evaluated as standalone components, yet real-world LLM deployments increasingly integrate retrieval-augmented generation with external knowledge stores and tool-integrated agent workflows. In RAG systems, sensitive information may be exposed through privacy inference over retrieved corpora and vector representations \cite{zeng2024good}, while tool-integrated agents remain vulnerable to indirect prompt injections embedded in untrusted external content \cite{zhan2024injecagent}. }
Effective deployment requires seamless integration of defenses into these pipelines. For example, training data cleaning must be embedded into data preprocessing workflows, while output \textcolor{black}{guardrails} should operate as a modular filter after response generation. \textcolor{black}{A practical example is training-data deduplication, which has been shown to reduce memorization and improve data quality in language model training pipelines \cite{lee2022deduplicating}. In web-scale data pipelines, hash-based integrity checks and source verification can further reduce split-view or frontrunning poisoning risks by verifying that crawled content remains consistent with the intended data source \cite{carlini2024poisoning}.}}

\textcolor{black}{Inference-time defenses, particularly output \textcolor{black}{guardrails} and adversarial detection, introduce additional processing latency. Real-time applications such as conversational agents may require responses within sub-second latency budgets. \textcolor{black}{For such deployment settings, modular guardrails such as WildGuard offer a practical example of input-output moderation, where prompts, responses, and refusal behavior are classified before unsafe content is returned to users \cite{han2024wildguard}.} Lightweight defense variants, such as specialized classifiers and selective output monitoring, offer practical alternatives for latency-sensitive deployments \cite{ghosh2025aegis2}. Many defenses evaluated on small-scale benchmarks (e.g., SST-2) may also fail to scale to industrial data volumes. Data sanitization techniques that require exhaustive anomaly detection become prohibitively expensive on terabyte-scale pretraining corpora. Future defense designs should prioritize sublinear or streaming algorithms suitable for large-scale data processing.}

\textcolor{black}{Deployed defenses must align with emerging regulations such as the EU AI Act, GDPR, and CCPA. For instance, the "right to be forgotten" imposes legal requirements for data deletion, making machine unlearning not only a technical challenge but also a compliance necessity \cite{zhang2025correcting}. Defenses should produce auditable evidence of compliance, including data provenance records \cite{kale2023provenance}. 
Regarding differential privacy (DP), the Laplace mechanism can achieve higher utility at moderate privacy budgets ($\varepsilon \approx 6-15$), particularly on high-dimensional datasets \cite{min2026varepsilon}. Under strong privacy protection ($\varepsilon < 3$), the Gaussian mechanism often causes significant performance degradation, with accuracy dropping to approximately 50\% on sentiment analysis tasks such as SST-2 \cite{yang2024lmo}. \textcolor{black}{For reproducibility, DP-based defenses should explicitly report the privacy budget $(\varepsilon,\delta)$, clipping norm, noise multiplier, noise distribution, and noise calibration method, since these parameters determine both the privacy guarantee and the resulting utility loss.} However, there is currently no widely accepted methodology for selecting $\varepsilon$, and the optimal privacy budget depends on the specific query, desired accuracy, and deployment context \cite{shafiq2025effectiveness}. Recent research has begun to explore a ``utility-first'' framework, where practitioners prioritize a desired level of utility and then determine the corresponding privacy cost \cite{ghazi2026private}. This shift acknowledges that the optimal $\varepsilon$ remains an open problem requiring application-specific investigation. Organizations must also balance security investments against operational costs. A layered defense combining lightweight input filtering and selective output monitoring may offer a pragmatic balance for resource-constrained deployments. Quantitative guidance on the marginal returns of additional defense layers remains an open research question.}

\begin{table*}[htbp]
\caption{\textcolor{black}{Comparative evaluation of defense strategies for LLM data security.}}
\label{tab:defense_comparison}
\centering
\color{black}
\footnotesize
\setlength{\tabcolsep}{4pt}
\renewcommand{\arraystretch}{1.10}
\begin{tabularx}{\textwidth}{@{}%
>{\RaggedRight\arraybackslash}p{0.110\linewidth}%
>{\RaggedRight\arraybackslash}p{0.120\linewidth}%
>{\RaggedRight\arraybackslash}p{0.230\linewidth}%
>{\centering\arraybackslash}m{0.060\linewidth}%
>{\centering\arraybackslash}m{0.050\linewidth}%
>{\centering\arraybackslash}m{0.070\linewidth}%
>{\centering\arraybackslash}m{0.070\linewidth}%
>{\RaggedRight\arraybackslash}X%
@{}}
\toprule
\makecell[tl]{\textbf{Defense}} &
\makecell[tl]{\textbf{Threat}\\\textbf{Type}} &
\makecell[tl]{\textbf{Robustness}\\\textbf{Metrics}} &
\makecell[c]{\textbf{Scale}} &
\makecell[c]{\textbf{Cost}} &
\makecell[c]{\textbf{FP/FN}} &
\makecell[c]{\textbf{RAG}\\\textbf{Agent}} &
\makecell[tl]{\textbf{Deployment}} \\
\midrule

\makecell[tl]{Training\\Data\\Cleaning} &
\makecell[tl]{Data\\Poisoning} &
\begin{minipage}[t]{\linewidth}
\raggedright
Med\\
(ASR $\downarrow$)\\
\scriptsize\cite{li2026backdoorllm}
\end{minipage} &
\makecell[c]{Yes} &
\makecell[c]{Low} &
\makecell[c]{L/M} &
\makecell[c]{Med} &
Pre-train / fine-tune pipelines \\[2pt]

\makecell[tl]{Adversarial\\Training} &
\makecell[tl]{Prompt\\Injection} &
\begin{minipage}[t]{\linewidth}
\raggedright
Med--High\\
(ASR $\downarrow$, Clean Acc $\downarrow$)\\
\scriptsize\cite{geiping2021doesn}
\end{minipage} &
\makecell[c]{No} &
\makecell[c]{High} &
\makecell[c]{M/M} &
\makecell[c]{Low} &
Controlled fine-tuning only \\[2pt]

\makecell[tl]{Output\\\textcolor{black}{Guardrails}} &
\makecell[tl]{Jailbreak} &
\begin{minipage}[t]{\linewidth}
\raggedright
Med\\
(ASR $\downarrow$)\\
\scriptsize\cite{han2024wildguard}
\end{minipage} &
\makecell[c]{Yes} &
\makecell[c]{Low} &
\makecell[c]{H/M} &
\makecell[c]{High} &
Black-box guardrails \\[2pt]

\makecell[tl]{RLHF} &
\makecell[tl]{Halluc-\\ination} &
\begin{minipage}[t]{\linewidth}
\raggedright
Med\\
(Halluc.\ $\downarrow$)\\
\scriptsize\cite{ouyang2022training}
\end{minipage} &
\makecell[c]{Part.} &
\makecell[c]{High} &
\makecell[c]{M/M} &
\makecell[c]{Low} &
Alignment; costly updates \\[2pt]

\makecell[tl]{Data\\Augmen-\\tation} &
\makecell[tl]{Bias} &
\begin{minipage}[t]{\linewidth}
\raggedright
Low\\
(SEAT, EED $\downarrow$)\\
\scriptsize\cite{yu2023mixup}
\end{minipage} &
\makecell[c]{Yes} &
\makecell[c]{Low} &
\makecell[c]{L/H} &
\makecell[c]{Low} &
Bias mitigation only \\[2pt]

\makecell[tl]{\textcolor{black}{RAG}\\\textcolor{black}{Corpus}\\\textcolor{black}{Controls}} &
\makecell[tl]{\textcolor{black}{Retrieval-}\\\textcolor{black}{stage}} &
\begin{minipage}[t]{\linewidth}
{\color{black}\raggedright
Qualitative\\
(corpus integrity)\\
\scriptsize\cite{zou2025poisonedrag}}
\end{minipage} &
\makecell[c]{\textcolor{black}{Yes}} &
\makecell[c]{\textcolor{black}{Med}} &
\makecell[c]{\textcolor{black}{M/M}} &
\makecell[c]{\textcolor{black}{High}} &
\textcolor{black}{Provenance, sanitization, authorization-aware retrieval} \\[2pt]

\makecell[tl]{\textcolor{black}{Agent}\\\textcolor{black}{Runtime}\\\textcolor{black}{Controls}} &
\makecell[tl]{\textcolor{black}{Agent-}\\\textcolor{black}{based}} &
\begin{minipage}[t]{\linewidth}
{\color{black}\raggedright
Qualitative\\
(tool-misuse $\downarrow$)\\
\scriptsize\cite{zhan2024injecagent}}
\end{minipage} &
\makecell[c]{\textcolor{black}{Part.}} &
\makecell[c]{\textcolor{black}{Med}} &
\makecell[c]{\textcolor{black}{L/M}} &
\makecell[c]{\textcolor{black}{High}} &
\textcolor{black}{Sandboxing, least privilege, memory quarantine} \\
\midrule
\multicolumn{8}{@{}p{\textwidth}@{}}{\scriptsize
ASR: Attack Success Rate; SEAT: Sentence Embedding Association Test; EED: Expected Equality Difference.
FP/FN: false-positive/false-negative trade-off (L/M/H = Low/Med/High).
RAG/Agent: relevance to RAG/agent deployments. Part.: Partial.
Deployment stage and model-access assumptions are discussed in Attack Assumptions.
$\downarrow$: lower is better.
\textcolor{black}{RAG Corpus Controls and Agent Runtime Controls are summarized qualitatively from deployment implications of the surveyed RAG/agent attacks, not from a unified quantitative protocol.}} \\
\bottomrule
\end{tabularx}
\color{black}
\end{table*}

\textcolor{black}{Table~\ref{tab:defense_comparison} should be viewed as an interpretive synthesis rather than a cross-study ranking. The underlying works differ in models, datasets, threat assumptions, attack budgets, and metrics, so qualitative levels indicate reported directional tendencies rather than comparable effect sizes. These summaries are based on multiple representative studies in our survey sample; model families and datasets vary (see the cited works).}
\textcolor{black}{Table~\ref{tab:defense_comparison} shows that output guardrails and training data cleaning are frequently reported as offering a favorable deployment trade-off for production LLM systems, with relatively low cost, strong scalability, and measurable ASR reduction under the examined settings.}
\textcolor{black}{Consistent with the threat categories in Table~\ref{tab:defense_comparison}, training data cleaning is primarily suited to data poisoning, adversarial training to prompt injection, output guardrails to jailbreak, RLHF to hallucination, and data augmentation to bias mitigation.}
\textcolor{black}{RAG corpus controls and agent runtime controls further address retrieval-stage and agent-based risks that the generic families cover only partially.}
\textcolor{black}{In deployment readiness terms, training data cleaning and output guardrails are already applicable in operational pipelines, while adversarial training and RLHF remain more resource-intensive and are often better suited to controlled fine-tuning or periodic model updates. Data augmentation remains useful for bias mitigation but offers limited coverage for other threat classes.}
\textcolor{black}{Regarding FP/FN trade-offs, output guardrails may increase false positives through over-refusal, whereas data augmentation may miss adaptive attacks; for RAG/agent settings, corpus-level cleaning and output guardrails are among the more relevant practical choices reported in our sample.}
\textcolor{black}{When retrieval corpora or tool-integrated agents are involved, these choices should be complemented by the deployment-specific RAG/agent controls in Table~\ref{tab:defense_comparison}.}
\textcolor{black}{These comparisons indicate that defense selection should be guided by the target threat type, robustness metric, scalability, deployment suitability, FP/FN tolerance, RAG/agent relevance, and cost rather than by robustness alone. We therefore recommend a layered defense combining cleaning, guardrails, and RLHF for high-stakes applications (e.g., healthcare, finance), and lightweight filtering for general-purpose deployments.}
\textcolor{black}{For RAG- and agent-enabled deployments, this layered design should additionally include provenance-aware corpus controls and runtime constraints on tools and memory.}

\begin{table*}[htbp]
\centering
\caption{Quantitative trend comparison of defense strategies.}
\label{tab:defense_quantitative}
\renewcommand{\arraystretch}{1.1}
\setlength{\tabcolsep}{2.5pt}
\footnotesize
\color{black}
\begin{tabular}{llll}
\toprule
\textbf{Defense Strategy} & \textbf{Threat Type} & \textbf{Key Metrics} & \textbf{Reported Trend} \\
\midrule
Training Data Cleaning & Data Poisoning & ASR, Clean Acc & ASR $\downarrow\downarrow$, Clean Acc $\approx$ \\
Adversarial Training & Prompt Injection & ASR, Robust Acc & ASR $\downarrow$, Clean Acc $\downarrow$ (trade-off) \\
Output \textcolor{black}{Guardrails} & Jailbreak & ASR, F1 & ASR $\downarrow\downarrow\downarrow$, F1 $\uparrow\uparrow$ \\
RLHF & Hallucination & Hallucination Rate, AUROC & Hallucination $\downarrow$, AUROC $\uparrow$ \\
Data Augmentation & Bias & SEAT, EED, F1 & Bias metrics $\downarrow\downarrow$, task performance $\approx$ \\
\midrule
\multicolumn{4}{p{\dimexpr\textwidth-2\tabcolsep}}{\scriptsize ASR: Attack Success Rate; SEAT: Sentence Embedding Association Test; EED: Expected Equality Difference.} \\
\multicolumn{4}{p{\dimexpr\textwidth-2\tabcolsep}}{\scriptsize $\downarrow$: reduction; $\downarrow\downarrow$: substantial reduction; $\downarrow\downarrow\downarrow$: most substantial reduction; $\uparrow$: improvement; $\approx$: preserved.} \\
\multicolumn{4}{p{\dimexpr\textwidth-2\tabcolsep}}{\scriptsize \textcolor{black}{Deployment-specific RAG corpus controls and agent runtime controls are discussed qualitatively in the text and are omitted here because comparable cross-study quantitative trends are not yet available under a shared protocol.}} \\
\bottomrule
\end{tabular}
\end{table*}

\textcolor{black}{Table~\ref{tab:defense_quantitative} synthesizes interpretive trends from the surveyed literature under heterogeneous settings. Trend arrows summarize reported directional effects rather than comparable effect sizes, because the underlying studies differ in models, datasets, threat assumptions, attack budgets, and metrics. These summaries are based on multiple representative studies in our survey sample; model families and datasets vary across the cited works.}
\textcolor{black}{For comparing different defense mechanisms, we recommend selecting evaluation metrics that align with the target threat type: ASR for security threats (e.g., data poisoning, prompt injection, jailbreak), clean/robust accuracy for utility trade-offs, hallucination rate and AUROC for hallucination mitigation, SEAT/EED for bias evaluation, and F1 with computational overhead for production readiness.}
\textcolor{black}{Within the surveyed studies, the synthesized trends show that output guardrails are often associated with comparatively large reported ASR reductions against jailbreak attacks (ASR $\downarrow\downarrow\downarrow$, F1 $\uparrow\uparrow$), training data cleaning is commonly linked to strong protection against data poisoning while preserving clean accuracy (ASR $\downarrow\downarrow$, Clean Acc $\approx$), adversarial training offers moderate robustness gain at the cost of clean accuracy (ASR $\downarrow$, Clean Acc $\downarrow$), RLHF is frequently reported to reduce hallucinations with improved AUROC, and data augmentation tends to mitigate social bias without degrading task performance (Bias $\downarrow\downarrow$, task performance $\approx$).}
\textcolor{black}{No single defense dominates all threat types; instead, the choice of defense should be guided by the specific threat model and deployment constraints.}
\textcolor{black}{For production systems facing diverse threats, a layered combination of training data cleaning and output guardrails is a commonly favored trade-off in our sample.}
\textcolor{black}{For RAG- and agent-enabled deployments, these generic defenses should be complemented by the corpus and runtime controls summarized in Table~\ref{tab:defense_comparison}.}
\textcolor{black}{From a deployment-readiness perspective, training data cleaning and output guardrails are currently the most production-feasible options, whereas adversarial training and RLHF remain comparatively resource-intensive and are often better suited to controlled updates rather than frequent online deployment.}

\textcolor{black}{For practitioners selecting defense mechanisms, we offer the following principles. First, align defense selection with the specific data threat model. Against training-time data threats (e.g., data poisoning), training data cleaning is preferred as it preserves clean accuracy while substantially reducing ASR.}
\textcolor{black}{Against inference-time data threats (e.g., jailbreak, prompt leakage), output guardrails often show comparatively strong reported ASR reduction and are readily deployable without model retraining \cite{han2024wildguard}.}
\textcolor{black}{Second, consider the data utility-security trade-off. Adversarial training improves robustness but degrades clean accuracy; RLHF reduces hallucination-related data risks but requires expensive human feedback. Third, adopt a layered defense strategy combining training data cleaning, output guardrails, and RLHF for comprehensive data protection across the LLM lifecycle.}

\textcolor{black}{For practitioners deploying LLMs in real-world applications, the choice between open-source and closed-source models introduces distinct data security considerations. Closed-source models accessed via APIs require data transmission to external servers, introducing risks such as data logging, unauthorized retention, and potential incorporation into training sets. This limits defense feasibility (e.g., training data cleaning requires model-level access) and increases reliance on provider-side protections. Open-source models allow local deployment, enabling organizations to maintain full control over sensitive data. However, open-source models introduce unique data security risks.}
\textcolor{black}{First, white-box access to model weights expands the attack surface for parameter-level misuse and insecure fine-tuning workflows.}
\textcolor{black}{Second, large-scale empirical studies have identified widespread security flaws in public-facing LLM deployments, including insecure protocols, poor TLS configurations, and unauthenticated access, highlighting significant systemic vulnerabilities \cite{hou2025unveiling}.}
\textcolor{black}{Third, untrusted fine-tuning pipelines may implant backdoors or otherwise compromise model integrity \cite{zhang2025careful}.}
\textcolor{black}{Therefore, for high-sensitivity data, local open-source deployment with self-managed security monitoring is recommended; for general-purpose tasks, closed-source APIs may suffice. A hybrid approach offers the best balance for many organizations.}

\section*{Datasets}

In addition to addressing model vulnerabilities such as bias, hallucination, and limited defense against novel threats, it is also critical to select appropriate datasets to evaluate the robustness and safety of LLMs under different application scenarios. In this section, datasets are categorized and reviewed based on their \textcolor{black}{application scenarios, threat types, and benchmark categories}. \textcolor{black}{Generic NLP testbeds are summarized in Table~\ref{tab:example5_generic}, and security-specific and bias benchmarks are summarized in Table~\ref{tab:example5_security}.} \textcolor{black}{Each table reports dataset size, threat type, benchmark category, security use, and benchmark notes, including popularity and known limitations, rather than only providing descriptive scenario information.} \textcolor{black}{This separation makes clear that many widely used corpora remain generic NLP resources reused for security experiments, whereas only a smaller set is designed for security-oriented or bias and fairness evaluation.} This overview will assist researchers in selecting suitable datasets for studying data security risks and defense strategies in LLMs \textcolor{black}{across data poisoning assessment, bias and fairness analysis, prompt leakage testing, prompt injection robustness evaluation, and related defense studies.}

\textbf{Movie.}  
Movie datasets are often used to evaluate vulnerabilities in LLMs, especially concerning sentiment analysis. The SST-2 dataset \cite{kurita2020weight}, \cite{li2024badedit}, \cite{sharma2024towards}, \cite{wan2023poisoning}, \cite{yan2024backdooring} \textcolor{black}{is derived from movie reviews with binary positive or negative sentiment labels; the original Stanford Sentiment Treebank contains 11,855 sentences, while the commonly used GLUE SST-2 benchmark contains 67,349 training sentences.} The simplicity of this dataset makes it a frequent target for attack experiments, which aim to inject backdoors and assess the trustworthiness of a model. Similarly, IMDb \cite{kurita2020weight}, \cite{wan2023poisoning}, with 50,000 reviews, provides a larger and more balanced set, often used to evaluate adversarial robustness. However, one of the challenges with movie datasets, like OpenSubtitles \cite{sharma2024towards}, which includes dialogues from movies and TV shows, is that the informal and diverse language structures introduce complexities when detecting adversarial manipulations. Rotten Tomatoes \cite{hui2024pleak}, \textcolor{black}{which contains 10,662 movie-review sentences with positive or negative labels,} brings forth concerns about prompt leakage and hallucination risks, where a model might generate incorrect or fabricated sentiments. The potential for biased or harmful outputs due to these vulnerabilities can compromise the reliability and credibility of LLMs, thereby emphasizing the importance of robust defense strategies.

\begin{table*}[htbp]
\tiny
\linespread{0.94}\selectfont
\centering
\color{black}
\caption{\textcolor{black}{Generic NLP Datasets.}}
\label{tab:example5_generic}
\renewcommand{\arraystretch}{1.32}
\setlength{\extrarowheight}{1pt}
\setlength{\baselineskip}{8.8pt}
\setlength{\tabcolsep}{4pt}
\begin{tabularx}{\textwidth}{@{}%
>{\RaggedRight\arraybackslash}p{0.065\linewidth}%
>{\RaggedRight\arraybackslash}p{0.110\linewidth}%
>{\RaggedRight\arraybackslash}p{0.088\linewidth}%
>{\RaggedRight\arraybackslash}p{0.088\linewidth}%
>{\RaggedRight\arraybackslash}p{0.112\linewidth}%
>{\RaggedRight\arraybackslash}p{0.122\linewidth}%
>{\RaggedRight\arraybackslash}X%
>{\RaggedRight\arraybackslash}p{0.065\linewidth}%
@{}}
\begin{minipage}[t]{\linewidth}\strut\raggedright\textbf{Scenario}\strut\end{minipage} &
\begin{minipage}[t]{\linewidth}\strut\raggedright\textbf{Dataset}\strut\end{minipage} &
\begin{minipage}[t]{\linewidth}\strut\raggedright\textbf{Size}\strut\end{minipage} &
\begin{minipage}[t]{\linewidth}\strut\raggedright\textcolor{black}{\textbf{Threat type}}\strut\end{minipage} &
\begin{minipage}[t]{\linewidth}\strut\raggedright\textcolor{black}{\textbf{Benchmark category}}\strut\end{minipage} &
\begin{minipage}[t]{\linewidth}\strut\raggedright\textbf{Security use}\strut\end{minipage} &
\begin{minipage}[t]{\linewidth}\strut\raggedright\textbf{Benchmark notes}\strut\end{minipage} &
\begin{minipage}[t]{\linewidth}\strut\raggedright\textbf{Ref}\strut\end{minipage} \\
\noalign{\vskip 1.5pt}\hline\noalign{\vskip 1.5pt}

 & \begin{minipage}[t]{\linewidth}\strut\raggedright SST-2\footnotemark[1]\strut\end{minipage} &
\begin{minipage}[t]{\linewidth}\strut\raggedright 67,349 training sentences\strut\end{minipage} &
\begin{minipage}[t]{\linewidth}\strut\raggedright \textcolor{black}{Data Poisoning}\strut\end{minipage} &
\begin{minipage}[t]{\linewidth}\strut\raggedright \textcolor{black}{Generic NLP / Sentiment}\strut\end{minipage} &
\begin{minipage}[t]{\linewidth}\strut\raggedright Sentiment poisoning and backdoor evaluation\strut\end{minipage} &
\begin{minipage}[t]{\linewidth}\strut\raggedright High popularity; binary sentiment labels and short inputs limit coverage of open-ended LLM security risks.\strut\end{minipage} &
\begin{minipage}[t]{\linewidth}\strut\raggedright\scriptsize \cite{kurita2020weight},\cite{li2024badedit}, \cite{sharma2024towards},\cite{wan2023poisoning}, \cite{yan2024backdooring}\strut\end{minipage} \\

 Movie & \begin{minipage}[t]{\linewidth}\strut\raggedright IMDb\footnotemark[2]\strut\end{minipage} &
\begin{minipage}[t]{\linewidth}\strut\raggedright 50,000 labeled reviews\strut\end{minipage} &
\begin{minipage}[t]{\linewidth}\strut\raggedright \textcolor{black}{Data Poisoning}\strut\end{minipage} &
\begin{minipage}[t]{\linewidth}\strut\raggedright \textcolor{black}{Generic NLP / Sentiment}\strut\end{minipage} &
\begin{minipage}[t]{\linewidth}\strut\raggedright Poisoning and instruction-tuning attack evaluation\strut\end{minipage} &
\begin{minipage}[t]{\linewidth}\strut\raggedright High popularity; narrow domain and sentiment labels are not security-specific.\strut\end{minipage} &
\begin{minipage}[t]{\linewidth}\strut\raggedright\scriptsize \cite{kurita2020weight},\cite{wan2023poisoning}\strut\end{minipage} \\

 & \begin{minipage}[t]{\linewidth}\strut\raggedright OpenSubtitles\footnotemark[3]\strut\end{minipage} &
\begin{minipage}[t]{\linewidth}\strut\raggedright Millions of subtitle pairs\strut\end{minipage} &
\begin{minipage}[t]{\linewidth}\strut\raggedright \textcolor{black}{Bias}\strut\end{minipage} &
\begin{minipage}[t]{\linewidth}\strut\raggedright \textcolor{black}{Generic NLP / Dialogue}\strut\end{minipage} &
\begin{minipage}[t]{\linewidth}\strut\raggedright Dialogue-based behavior and sycophancy evaluation\strut\end{minipage} &
\begin{minipage}[t]{\linewidth}\strut\raggedright Medium popularity; noisy subtitles and weak metadata reduce control over speaker intent and safety labels.\strut\end{minipage} &
\begin{minipage}[t]{\linewidth}\strut\raggedright\scriptsize \cite{sharma2024towards}\strut\end{minipage} \\

 & \begin{minipage}[t]{\linewidth}\strut\raggedright Rotten Tomatoes\footnotemark[4]\strut\end{minipage} &
\begin{minipage}[t]{\linewidth}\strut\raggedright 10,662 movie-review sentences\strut\end{minipage} &
\begin{minipage}[t]{\linewidth}\strut\raggedright \textcolor{black}{Prompt Leakage}\strut\end{minipage} &
\begin{minipage}[t]{\linewidth}\strut\raggedright \textcolor{black}{Generic NLP / Sentiment}\strut\end{minipage} &
\begin{minipage}[t]{\linewidth}\strut\raggedright Prompt leakage and black-box application evaluation\strut\end{minipage} &
\begin{minipage}[t]{\linewidth}\strut\raggedright Medium popularity; small scale and sentiment focus limit generalization to complex LLM applications.\strut\end{minipage} &
\begin{minipage}[t]{\linewidth}\strut\raggedright\scriptsize \cite{hui2024pleak}\strut\end{minipage} \\
\noalign{\vskip 1.5pt}\hline\noalign{\vskip 1.5pt}

 & \begin{minipage}[t]{\linewidth}\strut\raggedright AG News\footnotemark[5]\strut\end{minipage} &
\begin{minipage}[t]{\linewidth}\strut\raggedright 127,600 news samples\strut\end{minipage} &
\begin{minipage}[t]{\linewidth}\strut\raggedright \textcolor{black}{Data Poisoning}\strut\end{minipage} &
\begin{minipage}[t]{\linewidth}\strut\raggedright \textcolor{black}{Generic NLP / Topic\\classification}\strut\end{minipage} &
\begin{minipage}[t]{\linewidth}\strut\raggedright Topic-classification poisoning and backdoor evaluation\strut\end{minipage} &
\begin{minipage}[t]{\linewidth}\strut\raggedright High popularity; useful for controlled attacks but lacks native adversarial or safety annotations.\strut\end{minipage} &
\begin{minipage}[t]{\linewidth}\strut\raggedright\scriptsize \cite{kurita2020weight},\cite{li2024badedit}, \cite{wan2023poisoning},\cite{yan2024backdooring}\strut\end{minipage} \\

News & \begin{minipage}[t]{\linewidth}\strut\raggedright Financial\strut\end{minipage} &
\begin{minipage}[t]{\linewidth}\strut\raggedright 4,840 financial sentences\strut\end{minipage} &
\begin{minipage}[t]{\linewidth}\strut\raggedright \textcolor{black}{Prompt Leakage}\strut\end{minipage} &
\begin{minipage}[t]{\linewidth}\strut\raggedright \textcolor{black}{Domain NLP / Finance}\strut\end{minipage} &
\begin{minipage}[t]{\linewidth}\strut\raggedright Domain-specific prompt leakage or defense evaluation\strut\end{minipage} &
\begin{minipage}[t]{\linewidth}\strut\raggedright Medium popularity; small and sentiment-oriented, with limited coverage of broader financial security risks.\strut\end{minipage} &
\begin{minipage}[t]{\linewidth}\strut\raggedright\scriptsize \cite{hui2024pleak}\strut\end{minipage} \\

 & \begin{minipage}[t]{\linewidth}\strut\raggedright English Gigaword\strut\end{minipage} &
\begin{minipage}[t]{\linewidth}\strut\raggedright About 4M article-summary pairs\strut\end{minipage} &
\begin{minipage}[t]{\linewidth}\strut\raggedright \textcolor{black}{Prompt Injection}\strut\end{minipage} &
\begin{minipage}[t]{\linewidth}\strut\raggedright \textcolor{black}{Generic NLP / Summarization}\strut\end{minipage} &
\begin{minipage}[t]{\linewidth}\strut\raggedright Adversarial training and text-generation robustness evaluation\strut\end{minipage} &
\begin{minipage}[t]{\linewidth}\strut\raggedright High popularity; large but older news corpus, not designed specifically for LLM security benchmarking.\strut\end{minipage} &
\begin{minipage}[t]{\linewidth}\strut\raggedright\scriptsize \cite{wen2023adversarial}\strut\end{minipage} \\
\noalign{\vskip 1.5pt}\hline\noalign{\vskip 1.5pt}

Social & \begin{minipage}[t]{\linewidth}\strut\raggedright WikiText-2\strut\end{minipage} &
\begin{minipage}[t]{\linewidth}\strut\raggedright About 2.1M tokens\strut\end{minipage} &
\begin{minipage}[t]{\linewidth}\strut\raggedright \textcolor{black}{Bias}\strut\end{minipage} &
\begin{minipage}[t]{\linewidth}\strut\raggedright \textcolor{black}{Generic NLP / Language\\modeling}\strut\end{minipage} &
\begin{minipage}[t]{\linewidth}\strut\raggedright Language modeling and bias-related evaluation\strut\end{minipage} &
\begin{minipage}[t]{\linewidth}\strut\raggedright High popularity; general text corpus without explicit security or demographic annotations.\strut\end{minipage} &
\begin{minipage}[t]{\linewidth}\strut\raggedright\scriptsize \cite{sharma2024towards}\strut\end{minipage} \\
\noalign{\vskip 1.5pt}\hline\noalign{\vskip 1.5pt}

Book & \begin{minipage}[t]{\linewidth}\strut\raggedright BookCorpus\strut\end{minipage} &
\begin{minipage}[t]{\linewidth}\strut\raggedright 11,038 books\strut\end{minipage} &
\begin{minipage}[t]{\linewidth}\strut\raggedright \textcolor{black}{Data Poisoning}\strut\end{minipage} &
\begin{minipage}[t]{\linewidth}\strut\raggedright \textcolor{black}{Generic NLP / Pre-training\\corpus}\strut\end{minipage} &
\begin{minipage}[t]{\linewidth}\strut\raggedright Pre-training corpus and data contamination analysis\strut\end{minipage} &
\begin{minipage}[t]{\linewidth}\strut\raggedright High popularity; copyright, provenance, and duplication issues complicate security auditing.\strut\end{minipage} &
\begin{minipage}[t]{\linewidth}\strut\raggedright\scriptsize \cite{sharma2024towards},\cite{yu2023mixup}\strut\end{minipage} \\
\noalign{\vskip 1.5pt}\hline\noalign{\vskip 1.5pt}

Study & \begin{minipage}[t]{\linewidth}\strut\raggedright AQuA\footnotemark[6]\strut\end{minipage} &
\begin{minipage}[t]{\linewidth}\strut\raggedright 97,975 algebra questions\strut\end{minipage} &
\begin{minipage}[t]{\linewidth}\strut\raggedright \textcolor{black}{Prompt Injection}\strut\end{minipage} &
\begin{minipage}[t]{\linewidth}\strut\raggedright \textcolor{black}{Generic NLP / Reasoning QA}\strut\end{minipage} &
\begin{minipage}[t]{\linewidth}\strut\raggedright Reasoning attack and prompt manipulation evaluation\strut\end{minipage} &
\begin{minipage}[t]{\linewidth}\strut\raggedright Medium popularity; arithmetic reasoning focus limits coverage of broader LLM data-security threats.\strut\end{minipage} &
\begin{minipage}[t]{\linewidth}\strut\raggedright\scriptsize \cite{zhang2024goal}\strut\end{minipage} \\

 & \begin{minipage}[t]{\linewidth}\strut\raggedright TREC\strut\end{minipage} &
\begin{minipage}[t]{\linewidth}\strut\raggedright 5,952 labeled questions\strut\end{minipage} &
\begin{minipage}[t]{\linewidth}\strut\raggedright \textcolor{black}{Bias}\strut\end{minipage} &
\begin{minipage}[t]{\linewidth}\strut\raggedright \textcolor{black}{Generic NLP / Question\\classification}\strut\end{minipage} &
\begin{minipage}[t]{\linewidth}\strut\raggedright Data augmentation and robustness evaluation\strut\end{minipage} &
\begin{minipage}[t]{\linewidth}\strut\raggedright High popularity; small classification benchmark with limited coverage of modern LLM generation risks.\strut\end{minipage} &
\begin{minipage}[t]{\linewidth}\strut\raggedright\scriptsize \cite{yu2023mixup}\strut\end{minipage} \\
\noalign{\vskip 1.5pt}\hline\noalign{\vskip 1.5pt}
\end{tabularx}
\end{table*}

\begin{table*}[htbp]
\tiny
\linespread{0.94}\selectfont
\centering
\color{black}
\caption{\textcolor{black}{Security-specific and bias benchmarks.}}
\label{tab:example5_security}
\renewcommand{\arraystretch}{1.32}
\setlength{\extrarowheight}{1pt}
\setlength{\baselineskip}{8.8pt}
\setlength{\tabcolsep}{4pt}
\begin{tabularx}{\textwidth}{@{}%
>{\RaggedRight\arraybackslash}p{0.065\linewidth}%
>{\RaggedRight\arraybackslash}p{0.110\linewidth}%
>{\RaggedRight\arraybackslash}p{0.088\linewidth}%
>{\RaggedRight\arraybackslash}p{0.088\linewidth}%
>{\RaggedRight\arraybackslash}p{0.112\linewidth}%
>{\RaggedRight\arraybackslash}p{0.122\linewidth}%
>{\RaggedRight\arraybackslash}X%
>{\RaggedRight\arraybackslash}p{0.065\linewidth}%
@{}}
\begin{minipage}[t]{\linewidth}\strut\raggedright\textbf{Scenario}\strut\end{minipage} &
\begin{minipage}[t]{\linewidth}\strut\raggedright\textbf{Dataset}\strut\end{minipage} &
\begin{minipage}[t]{\linewidth}\strut\raggedright\textbf{Size}\strut\end{minipage} &
\begin{minipage}[t]{\linewidth}\strut\raggedright\textcolor{black}{\textbf{Threat type}}\strut\end{minipage} &
\begin{minipage}[t]{\linewidth}\strut\raggedright\textcolor{black}{\textbf{Benchmark category}}\strut\end{minipage} &
\begin{minipage}[t]{\linewidth}\strut\raggedright\textbf{Security use}\strut\end{minipage} &
\begin{minipage}[t]{\linewidth}\strut\raggedright\textbf{Benchmark notes}\strut\end{minipage} &
\begin{minipage}[t]{\linewidth}\strut\raggedright\textbf{Ref}\strut\end{minipage} \\
\noalign{\vskip 1.5pt}\hline\noalign{\vskip 1.5pt}

 & \begin{minipage}[t]{\linewidth}\strut\raggedright Sycophancy-\\eval\footnotemark[7]\strut\end{minipage} &
\begin{minipage}[t]{\linewidth}\strut\raggedright Fewer than 1,000 examples\strut\end{minipage} &
\begin{minipage}[t]{\linewidth}\strut\raggedright \textcolor{black}{Bias}\strut\end{minipage} &
\begin{minipage}[t]{\linewidth}\strut\raggedright \textcolor{black}{Security / Alignment\\behavior}\strut\end{minipage} &
\begin{minipage}[t]{\linewidth}\strut\raggedright Sycophancy and preference-following attack evaluation\strut\end{minipage} &
\begin{minipage}[t]{\linewidth}\strut\raggedright Niche popularity; small proof-of-concept benchmark with limited coverage of real-world user preferences.\strut\end{minipage} &
\begin{minipage}[t]{\linewidth}\strut\raggedright\scriptsize \cite{sharma2024towards}\strut\end{minipage} \\

Social & \begin{minipage}[t]{\linewidth}\strut\raggedright Bias-in-Bios\footnotemark[8]\strut\end{minipage} &
\begin{minipage}[t]{\linewidth}\strut\raggedright 397,340 biographies\strut\end{minipage} &
\begin{minipage}[t]{\linewidth}\strut\raggedright \textcolor{black}{Bias}\strut\end{minipage} &
\begin{minipage}[t]{\linewidth}\strut\raggedright \textcolor{black}{Bias / Fairness}\strut\end{minipage} &
\begin{minipage}[t]{\linewidth}\strut\raggedright Gender bias and debiasing evaluation\strut\end{minipage} &
\begin{minipage}[t]{\linewidth}\strut\raggedright High popularity; mainly covers binary gender and occupation labels, limiting intersectional bias analysis.\strut\end{minipage} &
\begin{minipage}[t]{\linewidth}\strut\raggedright\scriptsize \cite{gallegos2024bias}\strut\end{minipage} \\

 & \begin{minipage}[t]{\linewidth}\strut\raggedright Jigsaw\footnotemark[9]\strut\end{minipage} &
\begin{minipage}[t]{\linewidth}\strut\raggedright 159,571 labeled comments\strut\end{minipage} &
\begin{minipage}[t]{\linewidth}\strut\raggedright \textcolor{black}{Bias}\strut\end{minipage} &
\begin{minipage}[t]{\linewidth}\strut\raggedright \textcolor{black}{Bias / Fairness /\\Toxicity}\strut\end{minipage} &
\begin{minipage}[t]{\linewidth}\strut\raggedright Toxicity, bias, and counterfactual debiasing evaluation\strut\end{minipage} &
\begin{minipage}[t]{\linewidth}\strut\raggedright High popularity; labels may be noisy and reflect annotator subjectivity or platform-specific norms.\strut\end{minipage} &
\begin{minipage}[t]{\linewidth}\strut\raggedright\scriptsize \cite{wan2023poisoning},\cite{tokpo2024fairflow}, \cite{maudslay2019s}\strut\end{minipage} \\

 & \begin{minipage}[t]{\linewidth}\strut\raggedright ECHR\footnotemark[10]\strut\end{minipage} &
\begin{minipage}[t]{\linewidth}\strut\raggedright About 11,000 legal cases\strut\end{minipage} &
\begin{minipage}[t]{\linewidth}\strut\raggedright \textcolor{black}{Bias}\strut\end{minipage} &
\begin{minipage}[t]{\linewidth}\strut\raggedright \textcolor{black}{Bias / Fairness / Legal}\strut\end{minipage} &
\begin{minipage}[t]{\linewidth}\strut\raggedright Legal text robustness and fairness-related evaluation\strut\end{minipage} &
\begin{minipage}[t]{\linewidth}\strut\raggedright Medium popularity; domain-specific legal language limits transferability to general LLM security settings.\strut\end{minipage} &
\begin{minipage}[t]{\linewidth}\strut\raggedright\scriptsize \cite{gallegos2024bias}\strut\end{minipage} \\
\noalign{\vskip 1.5pt}\hline\noalign{\vskip 1.5pt}
\end{tabularx}
\end{table*}

\footnotetext[1]{\url{https://github.com/neulab/RIPPLe}}
\footnotetext[2]{\url{https://github.com/alexwan0/poisoning-instruction-tuned-models}}
\footnotetext[3]{\url{https://opus.nlpl.eu/OpenSubtitles/corpus/version/OpenSubtitles}}
\footnotetext[4]{\url{https://github.com/BHui97/PLeak}}
\footnotetext[5]{\url{https://github.com/wegodev2/virtual-prompt-injection}}
\footnotetext[6]{\url{https://worksheets.codalab.org/worksheets/0xbdd35bdd83b14f6287b24c9418983617/}}
\footnotetext[7]{\url{https://huggingface.co/datasets/meg-tong/sycophancy-eval}}
\footnotetext[8]{\url{https://github.com/i-gallegos/Fair-LLM-Benchmark}}
\footnotetext[9]{\url{https://github.com/rowanhm/counterfactual-data-substitution}}
\footnotetext[10]{\url{https://github.com/WenRuiUSTC/EntF}}

\textbf{News.}  
News datasets are indispensable for understanding the vulnerabilities and biases in LLMs, as they often serve as testing grounds for adversarial attacks and defense mechanisms. AG News \cite{kurita2020weight}, \cite{li2024badedit}, \cite{wan2023poisoning}, \cite{yan2024backdooring} \textcolor{black}{contains 127,600 news samples in total, including 120,000 training samples and 7,600 test samples,} categorized into World, Sports, Business, and Science. This variety makes it an ideal dataset for evaluating both attack models and the robustness of defenses. However, recent research points to the limitations of current defense strategies, as many are ineffective against new types of adversarial inputs. The Financial Dataset \cite{hui2024pleak}, \textcolor{black}{which contains 4,840 financial sentences from financial news, stock-market-related texts, and company reports,} presents unique challenges in domain-specific adversarial attacks, where subtle manipulations can cause significant errors in financial decision-making. English Gigaword \cite{wen2023adversarial}, \textcolor{black}{with about 4 million article-summary pairs from English news text,} highlights another issue: the difficulty in developing defense methods that can generalize well across various news categories and threat types. As we rely more and more on LLMs for real-world applications, ensuring their accuracy and reliability in these contexts becomes ever more critical.

\textbf{Social.}  
Social datasets reveal crucial challenges surrounding bias, fairness, and the ethical use of LLMs in sensitive areas such as legal and online social contexts. The Sycophancy-eval dataset \cite{sharma2024towards} \textcolor{black}{contains fewer than 1,000 examples and is used to evaluate sycophantic behavior in free-text generation,} a clear example of how the lack of control in conversational settings can result in undesirable model behavior. WikiText-2 \cite{sharma2024towards}, with about 2.1 million tokens from Wikipedia articles, also highlights the issue of biased content generation, as LLMs may perpetuate stereotypes or misinformation. Bias-in-Bios \cite{gallegos2024bias}, \textcolor{black}{which contains 397,340 biographies with occupation and gender labels,} raises ethical concerns about how models trained on biased data can reinforce societal inequalities. \textcolor{black}{Jigsaw \cite{wan2023poisoning}, \cite{tokpo2024fairflow}, \cite{maudslay2019s}, consisting of 159,571 labeled online comments, is widely used for toxicity, bias, and counterfactual debiasing evaluation.} \textcolor{black}{ECHR \cite{gallegos2024bias}, which contains about 11,000 legal cases from the European Court of Human Rights, supports legal text robustness and fairness-related evaluation.} These datasets collectively reflect a growing concern over how LLMs might exacerbate prejudices or unfair treatment, making it essential to develop more transparent and interpretable models.

\textbf{Book.}  
BookCorpus, a collection of \textcolor{black}{11,038 books} \cite{sharma2024towards}, \cite{yu2023mixup}, serves as a crucial resource for training LLMs. However, its complexity presents challenges in handling adversarial attacks, where subtle manipulations may lead to the generation of inaccurate or biased content. The vastness and diversity of the dataset increase the difficulty of maintaining context and factual accuracy in generated outputs. \textcolor{black}{Moreover, copyright, provenance, and duplication issues make BookCorpus a useful example for discussing data contamination and auditing limitations in large-scale pre-training corpora.} As a result, models trained on such large-scale datasets may struggle with hallucinations, creating information that does not exist. The need for transparency in these models becomes more apparent as understanding why certain content is generated is often difficult, leading to issues of trust and accountability in real-world applications \cite{bender2021dangers}.

\textbf{Study.}  
The AQuA dataset \cite{zhang2024goal}, \textcolor{black}{which contains 97,975 algebra questions with rationales,} is used to evaluate arithmetic problem-solving and highlights the challenges in ensuring precise reasoning in LLMs. Although it serves as a good benchmark for evaluating basic computational tasks, it exposes limitations in the abilities of models to generalize across diverse problem types, especially when faced with adversarial perturbations. Such weaknesses in defense mechanisms become particularly concerning in high-stakes applications where errors in calculations can have significant consequences. These challenges underscore the broader problem in the field: the need for more flexible, adaptive defense methods that can effectively handle novel threats and ensure the reliability and transparency of models in practical settings \cite{smiley2017say}, \cite{zhang2022improving}. The TREC dataset \cite{yu2023mixup}, \textcolor{black}{with 5,952 labeled questions,} evaluates information retrieval and question classification systems, supporting research into robust retrieval technologies and data augmentation.

\textbf{\textcolor{black}{Recommendations for Representative Benchmark Selection.}}  
\textcolor{black}{Based on our analysis, several datasets have emerged as de facto standards within specific threat categories. For data poisoning and backdoor attacks, SST-2 and AG News are most widely adopted due to their moderate scale and established evaluation protocols \cite{kurita2020weight, wan2023poisoning}. For prompt leakage and black-box application evaluation, Rotten Tomatoes and Financial datasets provide compact but domain-distinct testbeds. For bias evaluation, Jigsaw and Bias-in-Bios are widely used \cite{gallegos2024bias}. For reasoning and retrieval-related robustness, AQuA and TREC provide useful controlled benchmarks, although they remain limited in representing open-ended LLM generation risks. No single dataset suffices for comprehensive evaluation. A robust experimental design should combine datasets across multiple domains (movie, news, social, book, and study) and include both attack-oriented and defense-oriented purposes, drawing on the generic NLP testbeds in Table~\ref{tab:example5_generic} together with the security-specific and bias benchmarks in Table~\ref{tab:example5_security}.}
\textcolor{black}{As new threats targeting RAG systems emerge, existing datasets should be extended with security-specific annotations and retrieval-stage cases that reflect corpus integrity risks in external knowledge stores, such as poisoned or adversarially crafted documents \cite{zeng2024good}.
Benchmarks should also incorporate prompt-injected datastore manipulation scenarios to evaluate how malicious retrieved content can corrupt generation in retrieval-augmented pipelines \cite{qi2025follow}.
For LLM-based agents, evaluation resources should further include standardized agent-interaction traces and indirect prompt injection cases in tool-integrated workflows \cite{zhan2024injecagent}.}

\begin{table}[htbp]
\footnotesize
\centering
\color{black}
\caption{Metric glossary for LLM data-security evaluation.}
\label{tab:metric_glossary}
\renewcommand{\arraystretch}{1.12}
\setlength{\tabcolsep}{3.5pt}
\begin{tabularx}{\linewidth}{@{}%
>{\RaggedRight\arraybackslash}p{0.11\linewidth}%
>{\RaggedRight\arraybackslash}p{0.22\linewidth}%
>{\RaggedRight\arraybackslash}p{0.20\linewidth}%
>{\centering\arraybackslash}p{0.07\linewidth}%
>{\RaggedRight\arraybackslash}X%
@{}}
\toprule
\textbf{Metric} & \textbf{Definition} & \textbf{Computation} & \textbf{Better} & \textbf{Applicable threat(s)} \\
\midrule
ASR & Attack success rate & successes / attempts & $\downarrow$ & Poisoning, Injection, Jailbreak, Leakage \\

Clean Acc. & Benign-input accuracy & correct / benign & $\uparrow$ & Poisoning, Injection, Jailbreak \\

F1 & Detection balance score & $2PR/(P{+}R)$ & $\uparrow$ & Jailbreak, Injection \\

Halluc. Rate & Hallucinated output share & hallucinated / total & $\downarrow$ & Hallucination \\

AUROC & Hallucination discrimination & ROC-AUC & $\uparrow$ & Hallucination \\

SEAT & Embedding association bias & effect size in embedding space & $\downarrow$ & Bias \\

EED & Group outcome disparity & $\mathbb{E}[s|g_1]{-}\mathbb{E}[s|g_2]$ & $\downarrow$ & Bias \\

\textcolor{black}{Extr.\ Succ.} & \textcolor{black}{Extraction success} & \textcolor{black}{extracted / attempts} & \textcolor{black}{$\downarrow$} & \textcolor{black}{Extraction, Memorization, Privacy} \\

\textcolor{black}{Refusal} & \textcolor{black}{Unsafe-query refusal rate} & \textcolor{black}{refusals / unsafe queries} & \textcolor{black}{$\uparrow$} & \textcolor{black}{Jailbreak, Harmful prompts} \\

\textcolor{black}{Over-refusal} & \textcolor{black}{Benign-query false refusal} & \textcolor{black}{false refusals / benign} & \textcolor{black}{$\downarrow$} & \textcolor{black}{Guardrails, Utility trade-off} \\

\textcolor{black}{Harm.\ Comp.} & \textcolor{black}{Harmful-instruction compliance} & \textcolor{black}{complies / harmful queries} & \textcolor{black}{$\downarrow$} & \textcolor{black}{Jailbreak, Unsafe following} \\

\textcolor{black}{Poisoned-doc RR} & \textcolor{black}{Poisoned-document retrieval} & \textcolor{black}{poisoned / retrieved} & \textcolor{black}{$\downarrow$} & \textcolor{black}{RAG / Vector-store poisoning} \\

\textcolor{black}{Tool-misuse} & \textcolor{black}{Unsafe tool-use success} & \textcolor{black}{misuses / attempts} & \textcolor{black}{$\downarrow$} & \textcolor{black}{Agent / Tool attacks} \\

\textcolor{black}{Exfil.\ Succ.} & \textcolor{black}{Data-exfiltration success} & \textcolor{black}{exfiltrations / attempts} & \textcolor{black}{$\downarrow$} & \textcolor{black}{Prompt / RAG / Agent leakage} \\

\textcolor{black}{Latency / Cost} & \textcolor{black}{Defense runtime and cost} & \textcolor{black}{$\Delta$time, $\Delta$cost} & \textcolor{black}{$\downarrow$} & \textcolor{black}{Deployment-side defenses} \\

\textcolor{black}{DP $(\varepsilon,\delta)$} & \textcolor{black}{Privacy budget} & \textcolor{black}{reported $(\varepsilon,\delta)$} & \textcolor{black}{trade-off} & \textcolor{black}{DP training / Unlearning} \\

\textcolor{black}{Calibration} & \textcolor{black}{Confidence calibration (e.g., ECE)} & \textcolor{black}{ECE or related score} & \textcolor{black}{$\downarrow$} & \textcolor{black}{Hallucination, Uncertainty} \\
\midrule
\multicolumn{5}{p{\dimexpr\linewidth-2\tabcolsep}}{\scriptsize $\uparrow$ higher is better; $\downarrow$ lower is better.} \\
\multicolumn{5}{p{\dimexpr\linewidth-2\tabcolsep}}{\scriptsize ASR is the main attack metric; Clean Acc.\ and F1 capture security--utility trade-offs; SEAT and EED are bias metrics; Halluc.\ Rate and AUROC are used for hallucination evaluation.} \\
\multicolumn{5}{p{\dimexpr\linewidth-2\tabcolsep}}{\scriptsize \textcolor{black}{Security metrics are threat-specific and should not be compared across studies without aligned threat models, attack budgets, and evaluation protocols.}} \\
\bottomrule
\end{tabularx}
\color{black}
\end{table}

\textcolor{black}{To improve clarity across threat families, Table~\ref{tab:metric_glossary} summarizes the main evaluation metrics used in the surveyed LLM data-security literature. For each metric, we report its definition, computation, preferred direction, and applicable threat type. ASR is the primary attack-centric metric for data poisoning, prompt injection, jailbreak, and prompt leakage; Clean Accuracy and F1 are commonly used to assess security--utility trade-offs; Hallucination Rate and AUROC support hallucination evaluation; and SEAT and EED are standard bias-oriented metrics.} \textcolor{black}{We further include extraction success, refusal and over-refusal rates, harmful compliance, poisoned-document retrieval rate, tool-misuse success, data-exfiltration success, latency and cost overhead, differential-privacy parameters, and calibration metrics to cover privacy extraction, RAG/agent deployments, and deployment cost. Refusal on unsafe queries should be considered together with over-refusal on benign inputs, and differential-privacy budgets $(\varepsilon,\delta)$ should be reported alongside task utility because stronger privacy often reduces performance.} \textcolor{black}{Researchers should select metrics according to the threat under study rather than relying on accuracy alone.} \textcolor{black}{Because these metrics are threat-specific, numerical results should not be compared across studies unless the underlying threat models, datasets, and evaluation settings are aligned.}

\section*{Future Directions}
 
\subsection*{Data Provenance and Traceability}
The data sources of LLMs are extensive, involving multiple stages and participants. Ensuring the security of the entire supply chain, from data collection, storage to transmission and use, is crucial. It is necessary to establish data supply chain security standards and certification systems, conduct strict reviews of data suppliers, prevent malicious data injection or leakage, and ensure the integrity and availability of data. Apart from security issues, ensuring data provenance and traceability throughout the data pipeline is essential for model transparency and accountability. Recent work emphasizes that establishing machine-actionable provenance records helps build explainable and trustworthy AI systems by providing an auditable trail of how data influence model behavior \cite{kale2023provenance}.

In addition, traceability models and tools have been systematically reviewed as foundational components for ensuring the trustworthiness and reproducibility of AI systems, particularly under complex and heterogeneous data environments as seen in LLM development \cite{mora2021traceability}. Building on this, a comprehensive auditing framework has been proposed to close the AI accountability gap, highlighting the need to trace not only data inputs but also decision-making processes and model iterations across the entire development \cite{raji2020closing}. We should design systematic frameworks for tracking the origin, curation steps, and transformation history of every data-point used in LLM training. Existing studies have proposed a data management framework for responsible artificial intelligence, emphasizing the core role of data traceability in enhancing the transparency and compliance of models \cite{werder2022establishing}. Secure meta-data capture and verifiable audit trails will help to both attribute harmful model behaviors and facilitate responsible content sourcing.

\subsection*{Efficient and Verifiable Machine Forgetting}
\textcolor{black}{A critical vulnerability in deployed LLMs is their capacity to memorize and potentially reproduce sensitive, erroneous, or legally protected training data. Remedying such memorization currently presents a severe practical constraint: the only method with rigorous guarantees is complete model retraining, a prohibitively expensive and operationally burdensome process. While parameter editing and targeted fine-tuning offer more expedient alternatives, they lack formal theoretical foundations. These techniques cannot provide verifiable proof that the targeted data has been fully excised and often degrade model performance or inadvertently introduce new attack surfaces.}

\textcolor{black}{Consequently, the development of theoretically sound, efficient, and scalable machine unlearning algorithms represents a pressing research imperative. This challenge extends beyond pure algorithm design to encompass a broader systemic issue involving computational efficiency, formal verification, and privacy-preserving architectures. Current research is progressing along two primary axes. The first focuses on algorithmic mechanisms for precise data removal, leveraging influence functions \cite{zhang2025correcting} to estimate and invert the contribution of specific data points to model parameters. The second axis seeks to establish formal verification frameworks. By integrating principles from differential privacy, these frameworks aim to provide proofs that an unlearned model's behavior is statistically indistinguishable from that of a model never exposed to the removed data. Initial proposals, such as the federated unlearning scheme \cite{zhong2025hierarchical}, provide conceptual starting points. However, scaling these methods to accommodate the parameter space of modern LLMs, which routinely exceed hundreds of billions of parameters, without compromising practical efficiency or robust verifiability constitutes a major unresolved research frontier.}

\subsection*{Continual Learning for Secure Model Updates}
LLMs are incrementally updated with new data; therefore, there must be developed research mechanisms to ensure that each update cannot be exploited to inject backdoors or leak previously covered private information. Tracking cumulative privacy loss over multiple fine-tuning rounds will be essential.

Future work should investigate privacy-preserving continual learning frameworks that enable secure knowledge acquisition over time without exposing prior training data. \textcolor{black}{In continual learning settings, lifelong differential privacy provides a foundational framework that maintains performance across sequential tasks while bounding privacy loss over time and reducing risks of unintended cross-task information leakage, laying the groundwork for safer long-term model adaptation~\cite{lai2022lifelong}.} This is especially important as models interact with sensitive user inputs over time. While privacy concerns have been extensively discussed, data security risks - such as malicious prompt injection or the persistence of toxic content - remain under-addressed. LLMs can memorize and reproduce portions of their training data, which may include toxic or policy-violating content \cite{carlini2021extracting}. Meta-learning based continual learning approaches have been proposed to dynamically adjust model parameters during incremental updates, thereby improving resistance to adversarial attacks and reducing the risk of harmful behavior in LLMs \cite{li2017learning}. 

In spite of this, there is still a significant challenge to ensure that such harmful data does not degrade model behavior or introduce vulnerabilities over successive training rounds. The need for robust data curation processes, ongoing data sanitization, and rigorous security checks during model updates is necessary.
\textcolor{black}{Another emerging direction is security governance for autonomous AI agents that rely on persistent long-term memory and protocol-based tool ecosystems, including MCP-like (Model Context Protocol) architectures. In such systems, models continuously read and write external context across multi-step workflows, which increases the risk of indirect prompt injection in tool-integrated agents \cite{zhan2024injecagent}. Persistent memory further expands the attack surface by enabling memory poisoning and cross-session leakage, making memory lifecycle controls and auditable execution constraints a key future research priority \cite{chen2024agentpoison}.}
\subsection*{\textcolor{black}{Standardized Evaluation Framework}}

\textcolor{black}{Despite the proliferation of defense strategies for LLM data security, a critical gap remains. There is no standardized, comprehensive benchmarking framework that enables fair and reproducible comparison across different threat models and defense strategies. Current evaluations are fragmented, with different studies employing disparate datasets, attack configurations, evaluation metrics, and model architectures \cite{onitiju2026security}. 
This fragmentation makes it difficult to answer fundamental questions, such as which defense reduces ASR most effectively across diverse threat types and how a training data cleaning method compares to adversarial training under identical conditions.}

\textcolor{black}{Recent efforts have begun to address this gap. BackdoorLLM provides a unified benchmark for backdoor attacks and defenses, covering eight attack strategies across six model architectures and seven real world scenarios \cite{li2026backdoorllm}. The S\&S Benchmark integrates ten major risk categories and 76 fine grained risk points, achieving 98.6\% accuracy with a multi model judgment framework \cite{zhang2025safety}. SafeBench evaluates multimodal LLMs across 23 risk scenarios with 2,300 harmful query pairs on 24 models \cite{ying2026safebench}. Building on these advances, future work should develop a unified benchmarking framework with four core components: (i) a standardized threat model suite covering all threat categories in our data focused taxonomy (data poisoning, prompt injection, jailbreak, hallucination, prompt leakage, bias, \textcolor{black}{adaptive and multi-stage attacks, and RAG and agent security risks}); (ii) a multi dimensional evaluation protocol reporting security metrics (ASR, refusal rate), utility metrics (clean accuracy), and efficiency metrics (computational overhead); (iii) a reference model zoo spanning diverse scales (1B to 70B+) and architectures (LLaMA, GPT, Claude); and (iv) an automated open source evaluation pipeline to ensure reproducibility. Such a framework would enable rigorous, comparable, and actionable assessment of LLM data security defenses.}

\subsection*{Explainability-Driven Security Analysis}
Leverage interpretability tools (attention-flow analysis, saliency methods, concept activation vectors) are not just for model transparency but also for active defenses - e.g., detecting anomalous rationale patterns that signal poisoning, or flagging content that unduly reflects single training instances. It is crucial to focus on advancing these interpretability technologies in future research to create robust frameworks, which can enable real-time security monitoring of LLMs during incremental updates. For instance, anomaly detection methods have demonstrated potential in revealing unusual training patterns that may indicate adversarial manipulation \cite{paudice2018detection}. 
\textcolor{black}{Training data attribution methods highlight influential tokens and pretraining examples linked to a model generation, facilitating the discovery of suspicious outputs influenced by memorized or poisoned training data \cite{lin2024token}. } 
Integrating these tools into continual learning pipelines offers a promising direction to enhance the security and trustworthiness of LLMs as they evolve.

\subsection*{Ethical and Regulatory Frameworks for LLM Data Governance}
Because LLMs handle sensitive data globally, interdisciplinary efforts must define auditing standards, data sovereignty protocols, and liability frameworks. Collaboration with policymakers will ensure alignment with evolving regulations. Recent work has emphasized how global-scale models demand policy-aware oversight and formal responsibility allocation, especially when their decisions affect end users in high-stakes contexts \cite{bommasani2021opportunities}.

Moreover, bridge technical advances with policy are as follows: propose data-security standards and certifications for “safety-compliant” LLMs, inform privacy regulation (e.g., GDPR, CCPA) with concrete measurement methodologies, and develop governance models that enable redress when models inadvertently expose or misuse personal data. For such frameworks to become operational, future work should explore how system-level governance mechanisms can be embedded directly into the LLM development pipeline. An end-to-end internal algorithmic auditing framework - such as the one in the context of deployed AI systems - can inspire LLM-specific protocols that incorporate documentation, oversight checkpoints, and accountability mapping throughout the model lifecycle \cite{raji2020closing}. A further challenge is enabling user redress in cases where models inadvertently expose or misuse sensitive training data. To this end, governance models must incorporate mechanisms such as fine-grained data lineage tracking and post-hoc auditing of generation behavior. Embedding these governance principles into the training lifecycle itself, as suggested in recent work on the ethical risks of LLM deployment, may also enhance institutional trust and regulatory compliance \cite{weidinger2021ethical}.

\section*{Conclusion}
In this survey, we explored the critical issues surrounding data security risks in LLMs. As these models are increasingly deployed across a wide range of real-world applications, ensuring the integrity and safety of the data they consume during training and inference has become a pressing concern. We first discussed \textcolor{black}{eight} major types of data security risks, such as data poisoning, prompt injection, jailbreak, hallucination, prompt leakage, \textcolor{black}{bias, \textcolor{black}{adaptive and multi-stage attacks, and RAG and agent security risks}, which} may lead to harmful or manipulated outputs. We then reviewed several defense strategies, including adversarial training, training data cleaning, output guardrails, RLHF, data augmentation, and RAG/agent defenses, which can mitigate such threats by improving model robustness and trustworthiness. In addition, we presented a comparative analysis of existing datasets, categorized by domain, use cases (attack or defense), and key characteristics. Lastly, we proposed future research directions, including data provenance and traceability, efficient and verifiable machine forgetting, continual learning for secure model
updates, \textcolor{black}{standardized evaluation framework,} explainability-driven security analysis, and ethical and regulatory frameworks for LLM data governance. This systematic overview aims to empower researchers and practitioners in strengthening data security throughout the LLM lifecycle.


\section*{Ethics declarations}
This article does not contain studies with human participants or animals and does not involve identifiable personal data. Therefore, ethical approval and informed consent were not required.

\bibliography{SpringerBasic}  

\appendix

\section*{Appendix A: Full list of PRISMA-included studies}
\phantomsection\label{tab:prisma_appendix}
\textcolor{black}{List~A1 shows all 94 studies included through the PRISMA screening process.
Each entry is shown as \textbf{No.}, \textbf{Ref.}, and title, where \textbf{No.} is a sequential index (1--94)
and \textbf{Ref.} is the citation number in the reference list.}

\vspace{0.6em}
\noindent\textbf{List~A1. Full list of the 94 PRISMA-included studies.}\par
\vspace{0.5em}
\begingroup
\scriptsize
\raggedright
\setlength{\parindent}{0pt}
\noindent\textbf{1}\enspace\cite{nazary2025stealthy}\enspace Stealthy llm-driven data poisoning attacks against embedding-based retrieval-augmented recommender systems\par\smallskip
\noindent\textbf{2}\enspace\cite{wang2025your}\enspace Is your prompt safe? investigating prompt injection attacks against open-source llms\par\smallskip
\noindent\textbf{3}\enspace\cite{mkadry2017towards}\enspace Towards deep learning models resistant to adversarial attacks\par\smallskip
\noindent\textbf{4}\enspace\cite{ouyang2022training}\enspace Training language models to follow instructions with human feedback\par\smallskip
\noindent\textbf{5}\enspace\cite{zeng2024good}\enspace The good and the bad: Exploring privacy issues in retrieval-augmented generation (rag)\par\smallskip
\noindent\textbf{6}\enspace\cite{an2025rag}\enspace Rag llms are not safer: A safety analysis of retrieval-augmented generation for large language models\par\smallskip
\noindent\textbf{7}\enspace\cite{qi2025follow}\enspace Follow my instruction and spill the beans: Scalable data extraction from retrieval-augmented generation systems\par\smallskip
\noindent\textbf{8}\enspace\cite{zhan2024injecagent}\enspace Injecagent: Benchmarking indirect prompt injections in tool-integrated large language model agents\par\smallskip
\noindent\textbf{9}\enspace\cite{liu2020backdoor}\enspace Backdoor attacks and defenses in feature-partitioned collaborative learning\par\smallskip
\noindent\textbf{10}\enspace\cite{wang2025badlingual}\enspace Badlingual: A novel lingual-backdoor attack against large language models\par\smallskip
\noindent\textbf{11}\enspace\cite{kong2025wolf}\enspace Wolf hidden in sheep's conversations: Toward harmless data-based backdoor attacks for jailbreaking large language models\par\smallskip
\noindent\textbf{12}\enspace\cite{wan2023poisoning}\enspace Poisoning language models during instruction tuning\par\smallskip
\noindent\textbf{13}\enspace\cite{sivapiromrat2025multi}\enspace Multi-trigger poisoning amplifies backdoor vulnerabilities in llms\par\smallskip
\noindent\textbf{14}\enspace\cite{wallace2021concealed}\enspace Concealed data poisoning attacks on nlp models\par\smallskip
\noindent\textbf{15}\enspace\cite{kshetri2023cybercrime}\enspace Cybercrime and privacy threats of large language models\par\smallskip
\noindent\textbf{16}\enspace\cite{kurita2020weight}\enspace Weight poisoning attacks on pretrained models\par\smallskip
\noindent\textbf{17}\enspace\cite{he2025data}\enspace Data poisoning for in-context learning\par\smallskip
\noindent\textbf{18}\enspace\cite{chaudhari2025cascading}\enspace Cascading adversarial bias from injection to distillation in language models\par\smallskip
\noindent\textbf{19}\enspace\cite{llewellyn2025towards}\enspace Towards reliable and practical llm security evaluations via bayesian modelling\par\smallskip
\noindent\textbf{20}\enspace\cite{das2024exposing}\enspace Exposing vulnerabilities in clinical llms through data poisoning attacks: Case study in breast cancer\par\smallskip
\noindent\textbf{21}\enspace\cite{pathade2025red}\enspace Red teaming the mind of the machine: A systematic evaluation of prompt injection and jailbreak vulnerabilities in llms\par\smallskip
\noindent\textbf{22}\enspace\cite{alber2025medical}\enspace Medical large language models are vulnerable to data-poisoning attacks\par\smallskip
\noindent\textbf{23}\enspace\cite{hui2024pleak}\enspace Pleak: Prompt leaking attacks against large language model applications\par\smallskip
\noindent\textbf{24}\enspace\cite{greshake2023not}\enspace Not what you've signed up for: Compromising real-world llm-integrated applications with indirect prompt injection\par\smallskip
\noindent\textbf{25}\enspace\cite{zou2025poisonedrag}\enspace $\{$PoisonedRAG$\}$: Knowledge corruption attacks to $\{$Retrieval-Augmented$\}$ generation of large language models\par\smallskip
\noindent\textbf{26}\enspace\cite{li2024badedit}\enspace Badedit: Backdooring large language models by model editing\par\smallskip
\noindent\textbf{27}\enspace\cite{liu2023prompt}\enspace Prompt injection attack against llm-integrated applications\par\smallskip
\noindent\textbf{28}\enspace\cite{zhang2024goal}\enspace Goal-guided generative prompt injection attack on large language models\par\smallskip
\noindent\textbf{29}\enspace\cite{mckenzie2023inverse}\enspace Inverse scaling: When bigger isn't better\par\smallskip
\noindent\textbf{30}\enspace\cite{perez2022ignore}\enspace Ignore previous prompt: Attack techniques for language models\par\smallskip
\noindent\textbf{31}\enspace\cite{yan2024backdooring}\enspace Backdooring instruction-tuned large language models with virtual prompt injection\par\smallskip
\noindent\textbf{32}\enspace\cite{li2025llms}\enspace Llms caught in the crossfire: Malware requests and jailbreak challenges\par\smallskip
\noindent\textbf{33}\enspace\cite{shen2024anything}\enspace ' do anything now': Characterizing and evaluating in-the-wild jailbreak prompts on large language models\par\smallskip
\noindent\textbf{34}\enspace\cite{hackett2025bypassing}\enspace Bypassing llm guardrails: An empirical analysis of evasion attacks against prompt injection and jailbreak detection systems\par\smallskip
\noindent\textbf{35}\enspace\cite{cao2023autohall}\enspace Autohall: Automated hallucination dataset generation for large language models\par\smallskip
\noindent\textbf{36}\enspace\cite{jiang2024large}\enspace On large language models' hallucination with regard to known facts\par\smallskip
\noindent\textbf{37}\enspace\cite{farquhar2024detecting}\enspace Detecting hallucinations in large language models using semantic entropy\par\smallskip
\noindent\textbf{38}\enspace\cite{zou2023universal}\enspace Universal and transferable adversarial attacks on aligned language models\par\smallskip
\noindent\textbf{39}\enspace\cite{agarwal2024investigating}\enspace Investigating the prompt leakage effect and black-box defenses for multi-turn llm interactions\par\smallskip
\noindent\textbf{40}\enspace\cite{sharma2024towards}\enspace Towards understanding sycophancy in language models\par\smallskip
\noindent\textbf{41}\enspace\cite{liang2021towards}\enspace Towards understanding and mitigating social biases in language models\par\smallskip
\noindent\textbf{42}\enspace\cite{tramer2020adaptive}\enspace On adaptive attacks to adversarial example defenses\par\smallskip
\noindent\textbf{43}\enspace\cite{jiang2024turning}\enspace Turning generative models degenerate: The power of data poisoning attacks\par\smallskip
\noindent\textbf{44}\enspace\cite{liu2024formalizing}\enspace Formalizing and benchmarking prompt injection attacks and defenses\par\smallskip
\noindent\textbf{45}\enspace\cite{shi2024optimization}\enspace Optimization-based prompt injection attack to llm-as-a-judge\par\smallskip
\noindent\textbf{46}\enspace\cite{xu2024comprehensive}\enspace A comprehensive study of jailbreak attack versus defense for large language models\par\smallskip
\noindent\textbf{47}\enspace\cite{xu2024hallucination}\enspace Hallucination is inevitable: An innate limitation of large language models\par\smallskip
\noindent\textbf{48}\enspace\cite{zhang2025siren}\enspace Siren's song in the ai ocean: A survey on hallucination in large language models\par\smallskip
\noindent\textbf{49}\enspace\cite{li2023trustworthy}\enspace Trustworthy ai: From principles to practices\par\smallskip
\noindent\textbf{50}\enspace\cite{kalai2025language}\enspace Why language models hallucinate\par\smallskip
\noindent\textbf{51}\enspace\cite{omar2025multi}\enspace Multi-model assurance analysis showing large language models are highly vulnerable to adversarial hallucination attacks during clinical decision support\par\smallskip
\noindent\textbf{52}\enspace\cite{dodge2021documenting}\enspace Documenting large webtext corpora: A case study on the colossal clean crawled corpus\par\smallskip
\noindent\textbf{53}\enspace\cite{sheng2021societal}\enspace Societal biases in language generation: Progress and challenges\par\smallskip
\noindent\textbf{54}\enspace\cite{navigli2023biases}\enspace Biases in large language models: origins, inventory, and discussion\par\smallskip
\noindent\textbf{55}\enspace\cite{staab2024beyond}\enspace Beyond memorization: Violating privacy via inference with large language models\par\smallskip
\noindent\textbf{56}\enspace\cite{chen2024agentpoison}\enspace Agentpoison: Red-teaming llm agents via poisoning memory or knowledge bases\par\smallskip
\noindent\textbf{57}\enspace\cite{carlini2024poisoning}\enspace Poisoning web-scale training datasets is practical\par\smallskip
\noindent\textbf{58}\enspace\cite{li2026backdoorllm}\enspace Backdoorllm: A comprehensive benchmark for backdoor attacks and defenses on large language models\par\smallskip
\noindent\textbf{59}\enspace\cite{souly2025poisoning}\enspace Poisoning attacks on llms require a near-constant number of poison samples\par\smallskip
\noindent\textbf{60}\enspace\cite{fu2024poisonbench}\enspace Poisonbench: Assessing large language model vulnerability to data poisoning\par\smallskip
\noindent\textbf{61}\enspace\cite{han2024evaluation}\enspace An evaluation of the safety of chatgpt with malicious prompt injection\par\smallskip
\noindent\textbf{62}\enspace\cite{paudice2018detection}\enspace Detection of adversarial training examples in poisoning attacks through anomaly detection\par\smallskip
\noindent\textbf{63}\enspace\cite{geiping2021doesn}\enspace What doesn't kill you makes you robust (er): How to adversarially train against data poisoning\par\smallskip
\noindent\textbf{64}\enspace\cite{wen2023adversarial}\enspace Is adversarial training really a silver bullet for mitigating data poisoning?\par\smallskip
\noindent\textbf{65}\enspace\cite{han2024wildguard}\enspace Wildguard: Open one-stop moderation tools for safety risks, jailbreaks, and refusals of llms\par\smallskip
\noindent\textbf{66}\enspace\cite{zeng2024autodefense}\enspace Autodefense: Multi-agent llm defense against jailbreak attacks\par\smallskip
\noindent\textbf{67}\enspace\cite{ghosh2025aegis2}\enspace Aegis2. 0: A diverse ai safety dataset and risks taxonomy for alignment of llm guardrails\par\smallskip
\noindent\textbf{68}\enspace\cite{yu2024rlhf}\enspace Rlhf-v: Towards trustworthy mllms via behavior alignment from fine-grained correctional human feedback\par\smallskip
\noindent\textbf{69}\enspace\cite{burns2024discoveringlatentknowledgelanguage}\enspace Discovering Latent Knowledge in Language Models Without Supervision\par\smallskip
\noindent\textbf{70}\enspace\cite{tokpo2024fairflow}\enspace Fairflow: An automated approach to model-based counterfactual data augmentation for nlp\par\smallskip
\noindent\textbf{71}\enspace\cite{maudslay2019s}\enspace It's all in the name: Mitigating gender bias with name-based counterfactual data substitution\par\smallskip
\noindent\textbf{72}\enspace\cite{yu2023mixup}\enspace Mixup-based unified framework to overcome gender bias resurgence\par\smallskip
\noindent\textbf{73}\enspace\cite{zhou2025survey}\enspace A survey on backdoor threats in large language models (llms): Attacks, defenses, and evaluation methods\par\smallskip
\noindent\textbf{74}\enspace\cite{bhat2024puregen}\enspace Puregen: Universal data purification for train-time poison defense via generative model dynamics\par\smallskip
\noindent\textbf{75}\enspace\cite{sharma2025constitutional}\enspace Constitutional classifiers: Defending against universal jailbreaks across thousands of hours of red teaming\par\smallskip
\noindent\textbf{76}\enspace\cite{lee2022deduplicating}\enspace Deduplicating training data makes language models better\par\smallskip
\noindent\textbf{77}\enspace\cite{zhang2025correcting}\enspace Correcting large language model behavior via influence function\par\smallskip
\noindent\textbf{78}\enspace\cite{min2026varepsilon}\enspace The $\varepsilon$ dilemma balancing privacy and utility in differential privacy\par\smallskip
\noindent\textbf{79}\enspace\cite{yang2024lmo}\enspace Lmo-dp: optimizing the randomization mechanism for differentially private fine-tuning (large) language models\par\smallskip
\noindent\textbf{80}\enspace\cite{shafiq2025effectiveness}\enspace The effectiveness of differential privacy in real-world settings: A metrics-based framework to help practitioners visualise and evaluate $\varepsilon$\par\smallskip
\noindent\textbf{81}\enspace\cite{ghazi2026private}\enspace Private hyperparameter tuning with ex-post guarantee\par\smallskip
\noindent\textbf{82}\enspace\cite{carlini2019secret}\enspace The secret sharer: Evaluating and testing unintended memorization in neural networks\par\smallskip
\noindent\textbf{83}\enspace\cite{hou2025unveiling}\enspace Unveiling the landscape of llm deployment in the wild: An empirical study\par\smallskip
\noindent\textbf{84}\enspace\cite{zhang2025careful}\enspace Be careful when fine-tuning on open-source llms: Your fine-tuning data could be secretly stolen!\par\smallskip
\noindent\textbf{85}\enspace\cite{smiley2017say}\enspace Say the right thing right: Ethics issues in natural language generation systems\par\smallskip
\noindent\textbf{86}\enspace\cite{zhang2022improving}\enspace Improving the adversarial robustness of nlp models by information bottleneck\par\smallskip
\noindent\textbf{87}\enspace\cite{zhong2025hierarchical}\enspace Hierarchical federated unlearning for large language models\par\smallskip
\noindent\textbf{88}\enspace\cite{lai2022lifelong}\enspace Lifelong dp: Consistently bounded differential privacy in lifelong machine learning\par\smallskip
\noindent\textbf{89}\enspace\cite{carlini2021extracting}\enspace Extracting training data from large language models\par\smallskip
\noindent\textbf{90}\enspace\cite{li2017learning}\enspace Learning without forgetting\par\smallskip
\noindent\textbf{91}\enspace\cite{onitiju2026security}\enspace Security assessment and mitigation strategies for large language models: A comprehensive defensive framework\par\smallskip
\noindent\textbf{92}\enspace\cite{zhang2025safety}\enspace A safety and security-centered evaluation framework for large language models via multi-model judgment\par\smallskip
\noindent\textbf{93}\enspace\cite{ying2026safebench}\enspace Safebench: A safety evaluation framework for multimodal large language models\par\smallskip
\noindent\textbf{94}\enspace\cite{lin2024token}\enspace Token-wise influential training data retrieval for large language models\par\smallskip
\endgroup

\end{document}